%% file: iotj_jamming_underwater.tex
\documentclass[journal]{IEEEtran}
\ifCLASSINFOpdf
  \usepackage[pdftex]{graphicx}
\else
  \usepackage[dvips]{graphicx}
\fi
\usepackage{amsmath}
\usepackage{amsfonts}
\usepackage{amssymb}
\usepackage{algorithmic}

\usepackage{array}

\ifCLASSOPTIONcompsoc
  \usepackage[caption=false,font=normalsize,labelfont=sf,textfont=sf]{subfig}
\else
  \usepackage[caption=false,font=footnotesize]{subfig}
\fi

\usepackage{stfloats}

\usepackage{mathtools} 

\usepackage{makecell}

\usepackage[acronym]{glossaries}
\newacronym{usn}{USN}{Underwater acoustic Sensor Network}
\newacronym{dos}{DoS}{Denial-of-Service}
\newacronym{sinr}{SINR}{Signal to Interference and Noise Ratio}
\newacronym{snr}{SNR}{Signal to Noise Ratio}
\newacronym{fec}{FEC}{Forward Error Correction}
\newacronym{iot}{IoT}{Internet of Things}
\newacronym{pdr}{PDR}{Packet Delivery Ratio}
\newacronym{ne}{NE}{Nash Equilibrium}
\newacronym{bpsk}{BPSK}{Binary Phase Shift Keying}
\newacronym{crc}{CRC}{Cyclic Redundancy Check}
\newacronym{los}{LoS}{Line of Sight}
\newacronym{css}{CSS}{Chirp Spread Spectrum}
\newacronym{ber}{BER}{Bit Error Rate}
\newacronym{qpsk}{QPSK}{Quadrature Phase Shift Keying}
\newacronym{dqpsk}{DQPSK}{Differential Quadrature Phase Shift Keying}
\newacronym{rs}{RS}{Reed-Solomon}
\newacronym{fdma}{FDMA}{Frequency Division Multiple Access}

\newcommand{\prob}{{\rm P}}

\graphicspath{{./figures/}}
\usepackage{tikz}
\usetikzlibrary{arrows, decorations.markings}

\usepackage{pgfplots}
\pgfplotsset{compat=newest}
\usetikzlibrary{plotmarks}
\usetikzlibrary{arrows, decorations.markings,shapes,positioning,chains}
\usepgfplotslibrary{patchplots}
\usepackage{grffile}
\pgfplotsset{plot coordinates/math parser=false}
\newlength\fheight
\newlength\fwidth

\IEEEoverridecommandlockouts
\newcommand\copyrightnotice{%
\begin{tikzpicture}[remember picture,overlay]
\node[anchor=north,yshift=0pt] at (current page.north) {\fbox{\parbox{\dimexpr\textwidth-\fboxsep-\fboxrule\relax}{
\footnotesize \textcopyright 2020 IEEE. This paper has been accepted for presentation at IEEE IoTJ. Personal use of this material is permitted.
Permission from IEEE must be obtained for all other uses, in any current or future media,
including reprinting/republishing this material for advertising or promotional purposes,
creating new collective works, for resale or redistribution to servers or lists,
or reuse of any copyrighted component of this work in other works. DOI: 10.1109/JIOT.2020.2982613}}};
\end{tikzpicture}
}

\makeatletter
\newcommand\remembertext[2]{
  \immediate\write\@auxout{\unexpanded{\global\long\@namedef{mytext@#1}{\textit{#2}}}}%
  #2%
}
\newcommand\recalltext[1]{%
  \ifcsname mytext@#1\endcsname
    \@nameuse{mytext@#1}%
  \else
    ``??''
  \fi
}
\makeatother

\begin{document}
%
\title{A Game-Theoretic and Experimental Analysis of Energy-Depleting Underwater Jamming Attacks}
%
%
%

\author{Alberto Signori~\IEEEmembership{Student Member,~IEEE}, Federico Chiariotti~\IEEEmembership{Member,~IEEE}, Filippo Campagnaro~\IEEEmembership{Member,~IEEE}, Michele Zorzi~\IEEEmembership{Fellow,~IEEE}\vspace{-1cm}
\thanks{A. Signori (corresponding author, email: signoria@dei.unipd.it), F. Chiariotti, F. Campagnaro, and M. Zorzi are with the University of Padova, Department of Information Engineering, Padua, Italy.} 
\thanks{This paper is an extension of a previous work, presented at ACM WUWNet '19 with the title ``Jamming the underwater: a game-theoretic analysis of energy-depleting jamming attacks'' \cite{jamming_uw}.}

\thanks{Copyright (c) 2020 IEEE. Personal use of this material is permitted. However, permission to use this material for any other purposes must be obtained from the IEEE by sending a request to pubs-permissions@ieee.org.}
}

%
%

\markboth{IEEE Internet of Things Journal}%
{Signori \MakeLowercase{\textit{et al.}}: Underwater Jamming}
%



%
\maketitle

\copyrightnotice

\begin{abstract}
Security aspects in underwater wireless networks have not been widely investigated so far, despite the critical importance of the scenarios in which these networks can be employed. For example, an attack to a military underwater network for enemy targeting or identification can lead to serious consequences. Similarly, environmental monitoring applications such as tsunami prevention are also critical from a public safety point of view. 
In this work, we assess a scenario in which a malicious node tries to perform a jamming attack, degrading the communication quality of battery-powered underwater nodes. The legitimate transmitter may use packet level coding to increase the chances of correctly delivering packets. Because of the energy limitation of the nodes, the jammer's objective is twofold, namely: \emph{(i)} disrupting the communication and \emph{(ii)} reducing the lifetime of the victim by making it send more redundancy.  We model the jammer and the transmitter as players in a multistage game, deriving the optimal strategies. We evaluate the performance both in a model-based scenario and using real experimental data, and perform a sensitivity analysis to evaluate the performance of the strategies if the real channel model is different from the one they use.
\end{abstract}


\section{Introduction}\label{sec:intro}

\IEEEPARstart{U}{nderwater} sensor networks are enabling several military, industrial, and environmental applications: the ability to monitor the environment remotely is extremely useful for oil and gas platform and  pipeline maintenance, seabed erosion and tsunami risk mitigation, and coastal patrol. In this latter critical application, underwater sensor nodes can identify and target enemies, extending the monitoring range and allowing just a few patrol ships to cover a very wide area~\cite{kalwa_racun_project}.

However, underwater communications are hindered by the high attenuation of electromagnetic waves. For this reason, radio communications are only possible over very short-range broadband links~\cite{multimodal_wuwnet17}, and nodes at longer distances need to use acoustic waves: depending on the frequency, acoustic communication is possible at ranges from hundreds of meters up to tens of kilometers~\cite{Stojanovic}. However, even acoustic communications present some challenges, as sound waves have a low propagation speed, causing long delays, and the environmental noise caused by wind, marine life, and shipping activities can be very strong. Furthermore, there are strong multi-path effects due to the signal reflections with the bottom and the surface~\cite{UWsensor_Zorzi,Akyildiz05}.

In this already hostile environment, a \gls{dos} attack can be very effective, disabling the victim node's communications and disrupting the monitoring operation. The most effective way to perform it is by physical layer jamming: the attacker transmits a high-power signal, causing interference and blocking the correct reception of packets. There are effective countermeasures that  increase the robustness of the transmission to jamming attacks, such as power control and channel coding, but they come at a cost: the transmitter node needs to spend more energy to protect its transmissions, depleting its battery faster and reducing its lifetime.

In this work, we consider a jamming scenario in which the attacker has a double objective: it tries to block the legitimate communications, but it also turns the transmitter's own defense mechanism against it, forcing it to deplete its battery faster. We model the scenario using game theory, considering both nodes as rational players in a zero-sum multistage game, in which each burst of packets represents a subgame. The jammer decides how long it will jam the channel, while the legitimate transmitter chooses the amount of redundancy that it will add to each burst: both nodes are battery-limited, and they both  have a trade-off between increasing the probability of success in the current subgame and saving their battery.

We can derive the optimal long-term strategies for both players using a dynamic programming approach. For the sake of analytical tractability, in this work we focus on the case of \emph{complete information} available to both players, including the outcomes of each transmission attempt and the battery state of the nodes at any time. We study the trade-off between energy consumption and transmission success probability as a function of the distances between transmitter, receiver and jammer.

Our game-theoretic model and its Monte Carlo results were presented in~\cite{jamming_uw}. In the present work, we consider a more realistic setting considering the modulation used by S2C EvoLogics modems, and present the results obtained in a lake experiment with the aforementioned modems. Additionally, we expand the mathematical model by deriving 
recursive analytical formulations of the lifetime and success probability, using them to derive the results instead of Monte Carlo simulations. We also assess the computational complexity of the dynamic programming approach used to derive the optimal strategies for both players. Finally, we analyze what happens when the nodes do not have perfect information on the environment, and on the channel in particular, to test the robustness of the system in a real scenario where strategies are obtained in advance without knowing the actual channel conditions in the deployment location. We present the result of a sensitivity analysis on the packet error probabilities, as well as on the chosen channel model.

The rest of the paper is organized as follows. 
In Sec.~\ref{sec:related}, we give an overview of the state of the art models for \gls{dos} jamming attacks, focusing on the underwater scenario.
In Sec.~\ref{sec:model}, we present the game theoretic model and the system scenario. Sec.~\ref{sec:solution} explains how to derive the optimal strategies for the jammer and the transmitter. Sec.~\ref{sec:scenario} defines the evaluation scenario, Sec.~\ref{sec:results} describes the numerical evaluations, and Sec.~\ref{sec:conclusions} concludes the paper. 

\section{Related Work} \label{sec:related}
Physical layer jamming is one of the most common \gls{dos} attacks~\cite{petroccia_security}. The principle behind it is simple, yet powerful: a malicious node injects signals into the channel in order to deny or at least reduce service to the legitimate users by increasing their noise level and preventing them from receiving messages correctly. 
For instance, the attacker can send single-tone jamming signals or white Gaussian noise signals with the same bandwidth as the transmitter~\cite{jamming_comparison}. The latter approach makes the attacker more flexible, as the former is not effective if the transmitter uses spread-spectrum techniques, such as frequency hopping or direct sequence spread spectrum~\cite{domingo_security}. Other approaches, such as adaptive jamming, require the attacker to have an adaptive physical layer so as to change its modulation or transmission power. The victims of a jamming attack may passively adopt a simple duty cycling strategy~\cite{brownfield2005wireless}, or actively react to the \gls{dos} attack, e.g., by increasing their transmit power~\cite{chen2011fight} or using channel-hopping~\cite{alnifie2007multi}. In case of active defense, game theory is often used to model the interaction between the jammer and its victim: for instance, in~\cite{wang2018mitigating}, the theory is applied to a satellite transmission using frequency-hopping as a defense mechanism.
The main drawback of active defenses is typically an increased energy consumption, which means that the defenses themselves can be exploited by the attacker to deplete the victim node's battery and interrupt its transmissions. In this case, energy consumption needs to be included in the game formulation by introducing power constraints~\cite{mallik2000analysis}, or by considering nodes with limited energy. In the latter case, the jamming attack is typically modeled as a zero-sum game with a finite horizon~\cite{debruhl}, and  optimal strategies are derived by applying dynamic programming bottom-up, i.e., starting from the lowest energy levels and exploiting the solution to find the optimal strategy for higher energy levels. Our previous work~\cite{chiariotti2019game} applies this principle in an \gls{iot} network, exploiting retransmissions as part of the defense strategy. It also includes an analysis of the game with incomplete information on the jammer's capabilities.
 

Some recent works have also analyzed the jamming issue in the context of underwater acoustic networks.
For example, \cite{xiao2019reinforcement} applies a reinforcement learning deep Q-network-based transmission scheme, using movement as a countermeasure against a jamming attack in a mobile underwater acoustic network. The jammer sends acoustic signals with the same band as the transmitter, and each agent can decide its own transmission power level. The problem is modeled as a dynamic game in which all nodes are power-constrained; the winner of the game is the last node to completely deplete its battery. The results are proven via both simulation and a pool test, in short range.

\begin{figure}[t]
\centering
\includegraphics[width=0.88\columnwidth]{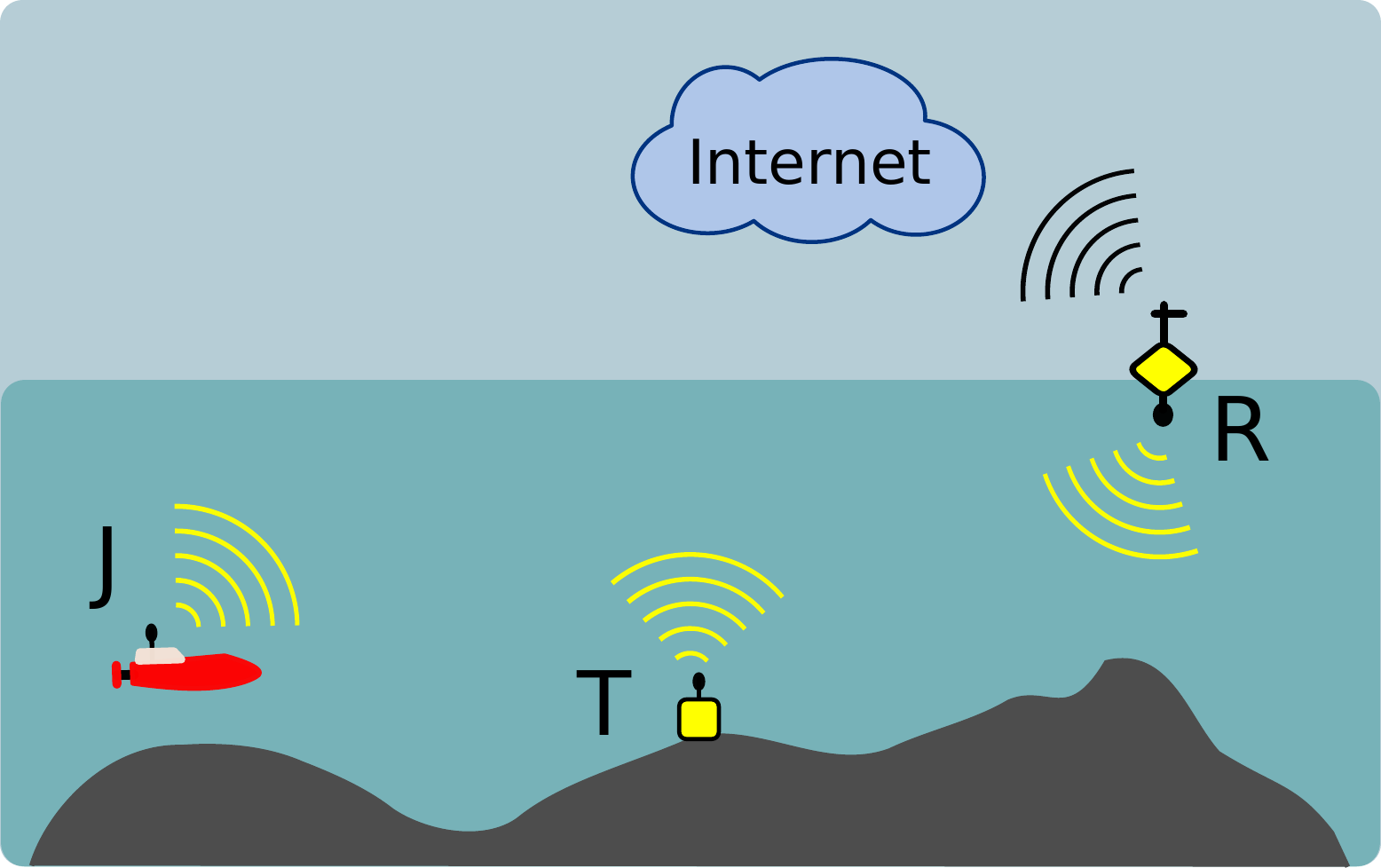}
\caption{An underwater jamming attack: a jammer $J$ tries disrupting the communication between a transmitter $T$ and its intended receiver $R$.} \label{fig:scenario}
\label{fig:model1}
\vspace{-3mm}
\end{figure}

In underwater acoustic networks, the propagation delay can be longer than the signal duration~\cite{HeidemannLasVegas}, especially in long range scenarios. In this case, a malicious node that observes the transmitter behavior and generates jamming signals as soon as it detects a new transmission cannot jam the current packet, since the jamming signal would reach the receiver only after the complete reception of the transmitted packet. Therefore, a jamming attack is effective in scenarios where the jamming signal reaches the receiver before the payload packet is completely received, such as when the jammer is placed between transmitter and receiver, or when the transmitter sends a sequence of packets with a deterministic or predictable pattern, such as in data muling applications.
In~\cite{block_jamming}, the authors propose a jamming defense strategy to provide secrecy for block transmission in underwater acoustic networks. They exploit the half-duplex nature of underwater transceivers and the large propagation delays to create interference at the eavesdropper. Specifically, the receiver transmits jamming packets to the malicious node during the guard time between data blocks, keeping the jammer transducer in the reception state and thereby preventing it from transmitting malicious signals. These packets do not cause deafness at the transmitter, as the propagation delay is larger than the guard time between blocks.
In \cite{vadori2015jamming}, a game-theoretic approach is applied to study jamming attacks in an underwater sensor network. Specifically, the authors analyzed a scenario in which two sensor nodes transmit their data to a sink node using \gls{fdma}. In this scenario, a jammer can disrupt the communication by injecting a signal only in one of the sub-bands used in the network. A Bayesian zero-sum game is considered to take into account the uncertainty on the distance between the sensor nodes and the sink. Differently from our work, the authors do not consider any power constraint in the problem formulation nor any form of active defense, such as packet level coding, and did not perform any field test to characterize how the presence of the jammer affects the reception.

\section{Game Theoretic Model}\label{sec:model}


We consider a transmitter $T$ at a distance $d_{\rm TR}$ from a receiver $R$, under attack from a jammer $J$, which is placed at a distance $d_{\rm JR}$ from $R$. The scenario is shown in Fig.~\ref{fig:scenario}: $T$ needs to periodically send an update to a receiver $R$, and a malicious jammer $J$ tries to block its transmission and deplete its battery. 

In order to protect its transmission from the attack and from ambient noise, $T$ uses packet-level coding: whenever it needs to send a message to $R$, it also sends a number of redundancy packets to protect the transmission from jamming attacks. Assuming an efficient packet-level code, the $K$ information packets can be recovered if at least any $K$ of the $N$ coded packets are correctly received~\cite{casari-rossi-08}.

The jamming attack is modeled as a zero-sum game $\mathbb{G}$ between the two rational players $T$ and $J$, i.e., a completely adversarial and symmetric game in which each gain for one player is balanced by a loss for the other~\cite{chiariotti2019game}. The zero-sum model is justified by the fact that an attacker will naturally want to disrupt the operation of the legitimate node as much as possible, thus having completely adversarial goals; this assumption is often used in jamming games. In this work, we study a complete information scenario. The complete information assumption is motivated by the fact that, at the end of each time frame, $R$ sends a feedback packet containing information on how many packets from $T$ it detected, how many slots were jammed by $J$, and how many packets it received successfully. We assume that such feedback packets are perfectly received by both $J$ and $T$, as $R$ is not power constrained.

The jamming game is composed of a series of packet transmission subgames $G_m$, with $m\in \mathbb{N}$. In each subgame, node $T$ uploads its data to node $R$, in an attempt to report information on the surrounding environment. Such data is chunked into $K$ payload packets, and $T$ can exploit \emph{(i)} \gls{fec} in order to increase the probability of successful communication over unreliable or noisy communication channels, and \emph{(ii)} \gls{crc} to detect residual error-laden packets and discard them. In each subgame $G_m$, $T$ can decide the amount of redundancy to use, i.e., the number $N_{T}^{(m)}$ of packets to send over the channel. A maximum of $2K$ transmission opportunities is configured in each subgame, thus $K \leq N_{T}^{(m)}\leq 2K$.

The outcome of each transmission attempt depends on the choices made by $T$ and $J$, and on the conditions of the channel, which is modeled stochastically. In particular, the transmission succeeds if $T$ is able to counteract the channel impairments \emph{and} the jamming attacks and to deliver at least $K$ packets to the destination node within the duration of the subgame. We assume a packet erasure channel and an efficient code, so $R$ can recover the $K$ information packets if any $K$ of the $N_T^{(m)}$ coded packets are correctly received~\cite{casari-rossi-08}. 
 
Both players are battery-powered nodes, and the dynamics of the game are exhaustively characterized by their energy evolution, i.e., the evolution of their battery charge during the game. The battery levels take discrete values in the sets $\mathcal{B}_i\triangleq [0,1,\dots,B_{i}^{(0)}],\,i\in\{T,J\}$, with $B_{i}^{(0)}\in\mathbb{N}$ being the initial charge of the battery.
The battery levels in the sets $\mathcal{B}_i$ are normalized by the energy $E_{{\rm tx},i}, \,i\in\{T,J\}$, used to transmit each legitimate packet or jam each slot; we consider the quantum $E_{{\rm tx},i}$ to be constant, since our active defense strategy does not involve power control. 
Note that, as neither energy harvesting nor other forms of energy replenishment are considered, the battery levels can only decrease during the game.  
In each subgame, node $T$ decides the number of packets $N_{T}^{(m)}$ to send to complete the data transmission, and this corresponds to an energy consumption of $N_{T}^{(m)}$ quanta, since battery levels are normalized.
Note that, the larger $N_{T}^{(m)}$, the more robust the communication, but the faster the depletion of $T$'s battery and the whole game duration.
Similar energy considerations affect the choice of the jammer, which has to decide the number of transmission opportunities $N_{J}^{(m)}$ to jam in order to disrupt $T$'s communication.

We now describe the structure of a single subgame and then illustrate the evolution of the multistage full game. Table~\ref{tab:notation} reports a summary of the notation used.

\begin{table}[t!]
\caption{Notation and meaning of system parameters for game players $i \in \{T,J\}$.} 
\vspace{-0.33cm}
\footnotesize
\centering
\begin{tabular}{cl}
\Xhline{2\arrayrulewidth}
Parameter & Meaning \\ \Xhline{2\arrayrulewidth}
$K$ & Minimum number of packets to be delivered for success \\
$\tau$ & Duration of a packet transmission \\
$\Gamma$  & Time horizon of multistage game $\mathbb{G}$  \\ 
$\lambda$  & Exponential discounting factor  \\ 
$\alpha_i$  & Energy/PDR weighing factor \\
$u_i^{(m)}$  & Payoff function in subgame $m$  \\ 
$U_i^{(m)}$  & Payoff function in multistage game $\mathbb{G}$ in subgame $m$  \\ 
$\chi_i^{(m)}$   & Indicator function of the success of subgame $m$ \\
$f_i^{(m)}$ & Energy penalty function  in subgame $m$ \\
$N_{T}^{(m)}$ & Number of packets that $T$ sends in subgame $m$\\
$N_{\rm C}^{(m)}$ & Packets sent over clear channel  in subgame $m$ \\
$N_{\rm B}^{(m)}$ & Packets sent over jammed channel  in subgame $m$ \\
$N_{J}^{(m)}$ & Number of slots that $J$ tries to jam in subgame $m$\\
$D^{(m)}$ & Total packets delivered  in subgame $m$\\
$d_{\rm C}^{(m)}$ & Packets delivered  over clear channel in subgame $m$\\
$p_{e_{\rm C}}$ & Packet error probability over clear channel \\
$p_{e_{\rm B}}$ & Packet error probability over jammed channel \\
$B_i^{(m)}$ & Battery level in subgame $m$ \\
$E_{{\rm tx},i}$ & Energy required to transmit/jam a packet \\
$P_{{\rm tx},i}$ & Transmission/jamming power \\
\Xhline{2\arrayrulewidth}
\end{tabular}
\label{tab:notation}
\vspace*{-0.1cm}
\end{table}

\subsection{The Packet Transmission Subgame} 
\label{sec:the_packet_transmission_subgame}

Each subgame $G_m$ models the attempt made by $T$ to transmit $K$ information packets to $R$.
The time after the beginning of the first packet transmission is slotted into a time frame of $2K$ time slots; each slot corresponds to the time $\tau$ necessary to transmit a packet.
Note that the long propagation delays that characterize the underwater scenario give an advantage to $T$: the first packet can never be jammed, as the jammer does not have the time to sense the transmission and send the jamming signal. However, since $J$ knows the duration of the time slot and the position of the transmitter and receiver, we assume that it can  trigger its transmissions to perfectly jam the subsequent time slots. 

Thus, $T$ decides \emph{(i)} how many packets $N_{T}^{(m)}\in\mathcal{N}_T^{(m)}\triangleq\{K,K+1,\dots,\min(2K,B_T^{(m)})\}$ to send to $R$, and \emph{(ii)} which time slots to employ for the transmission among the $2K$ available. Similarly, $J$ chooses \emph{(i)} the number of slots $N_{J}^{(m)}\in\mathcal{N}_J^{(m)}\triangleq\{0,1,\dots,\min(2K-1,B_J^{(m)})\}$ to jam, and \emph{(ii)} the $N_{J}^{(m)}$ jammed time slots out of $2K-1$ (as the first packet cannot be jammed). Note that the actions of both players are limited by the current battery level at stage $m$, i.e., $B_i^{(m)}$, $i\in\{T,J\}$. $T$ and $J$ make independent decisions on $N_{T}^{(m)}$ and $N_{J}^{(m)}$, respectively. Such decisions are made in advance for the whole time frame, right before the transmission of the first packet. 

The  payoffs of the players are convex combinations of monotonic functions of the energy required to transmit/jam the packets and of the \gls{pdr}.
By tuning the weight $\alpha\in[0,1]$, the main objective of the players can be shifted between saving energy, thereby reducing $N_{T}^{(m)}$ and $N_{J}^{(m)}$, and delivering more packets. 
Based on these considerations, we express the players' payoffs for a single subgame $m$ as:
\begin{align}
    u_{T}^{(m)} &=  \alpha \, f_T^{(m)} + (1-\alpha)\chi_T^{(m)}   \label{eq:payoffT} \\
    u_{J}^{(m)} &=  -u_T^{(m)}  \,.  \label{eq:payoffJ}
\end{align}
The first term of Eq.~\eqref{eq:payoffT} is related to energy, while the second term concerns the outcome of the communication. In particular, the indicator term $\chi_T^{(m)}$ is equal to one if the subgame $m$ ends with $T$ successfully delivering at least $K$ packets to $R$, and zero otherwise.

Function $f_T^{(m)}$ gives $T$ a penalty for consuming energy when transmitting packets. In particular, we set:
\begin{equation}
    f_T^{(m)} = -\frac{N_T^{(m)} }{(2K+1)}.
    \label{eq:f}
\end{equation}
The additional term 1 in the denominator of~\eqref{eq:f} is arbitrary and ensures that the absolute value of $f_T^{(m)}$ is always smaller than 1, thus preventing  any strategy to be dominated by not transmitting at all. 
Moreover, notice that the number of slots $N_J^{(m)}$ jammed by node $J$ is not explicitly present in the payoffs for the single subgame, since we assumed a zero-sum game. 
Nevertheless, $N_J^{(m)}$ still plays a major role in the full game: the larger $N_J^{(m)}$, the higher the energy consumed by node $J$, and the faster its battery depletion.

Finally, the transmitter's choice of the time slots in which to transmit packets, and the jammer's choice of which time slots to jam, can be modeled as a simple anti-coordination game: $T$'s objective is to avoid the jammer and transmit as many of its packets as possible on a clear channel, while $J$'s objective is to correctly guess the slots that $T$ will use and jam them, so as to maximally disrupt the communication.

\subsection{The Full Jamming Game} \label{sec:the_full_jamming_game}

In a battery-limited scenario, the greedy strategy that maximizes the payoff for the next subgame is not always optimal. 
The  solution of the full jamming game $\mathbb{G}$ maximizes a long-term payoff function within a given time horizon $\Gamma$, which represents the number of future subgames to consider in the payoff. 
The players' payoffs in the multistage game $\mathbb{G}$ at stage $m$ are given by:
\begin{equation}
    U_i^{(m)}(\Gamma)=\sum_{\gamma=m}^{m+\Gamma-1}\lambda^{\gamma-m}u_i^{(\gamma)},\quad i\in\{T,J\}\label{eq:fullpayoff}
\end{equation}
where $\lambda  \in [0, 1]$ is a future exponential discounting factor~\cite{abreu1988theory}, $u_i^{(m)}$, $i\in\{T,J\}$ is the subgame payoff defined in \eqref{eq:payoffT} and~\eqref{eq:payoffJ}, and $\Gamma$ is the length of the payoff horizon, i.e., the number of subgames that are considered. 
When $\Gamma$ is finite, we can consider $\lambda=1$ with no convergence issues, while, for $\Gamma=+\infty$, we must consider $\lambda<1$. Note that the payoff $u_i^{(m)}$ for a single subgame coincides with $U_i^{(m)}(1)$. In general, $J$ will behave in a foresighted manner if $\Gamma$ is large enough: its energy expenditure is not explicitly penalized, but it can reduce the reward if it affects the number of subgames it can play in.

\section{Analytical Solution of the Game}\label{sec:solution}

In this section, we explain how to derive the optimal strategies for the two players in the case of perfect knowledge about the opponent's position and battery level at the beginning of each subgame.
We define as strategy $s_i$ the action chosen by player $i\in\{T,J\}$, i.e., the amount of energy required to transmit the legitimate packets or to jam the slots, respectively. According to the game defined in Sec.~\ref{sec:model}, the strategy space is thus $\mathcal{N}_i^{(m)}\,i\in\{T,J\}$ in each subgame. 
Note that the strategies concern what to do in each subgame, but are chosen based on the expected evolution over multiple subgames, as dictated by $\Gamma$.
We are interested in evaluating the \gls{ne}, i.e., the pair of optimal strategies $(s_T^*,s_J^*)$ that are mutual best responses~\cite{nash}. In other words, a \gls{ne} is reached when neither player can improve its expected payoff by changing its strategy unilaterally.
Since the payoff functions of the two players (see~\eqref{eq:fullpayoff}) can include multiple subgames, the \gls{ne} of the jamming game can be calculated exactly with \emph{dynamic programming}.
The \gls{ne} may be \emph{pure}, i.e., correspond to deterministic strategies, or \emph{mixed}, when strategy $s_i^{(m)}$ for player $i\in\{T,J\}$ is a probability distribution $\Phi_{s_i}\left(N_i\right)$ over $\mathcal{N}_i^{(m)}$.
Under the assumption of complete information, strategies are determined by the state of the two players, assuming an optimal strategy for lower battery states. 

In the following, we first present the expressions for the expected payoffs of nodes $T$ and $J$ that are needed to compute the \gls{ne}, and then describe the procedure to solve the game analytically through dynamic programming.

\subsection{Expected Payoff Calculation}
\label{sec:exp_payoff}

To derive the \gls{ne}, we need to characterize the expected payoff for a single subgame, denoted as $\mathbb{E}\Big[U_{i}^{(m)}(1) \Big| N_T^{(m)},\,N_J^{(m)}\Big]$ for the $m$-th stage of game $\mathbb{G}$. Such expected payoff is equal to the expectation of the payoffs $u_i^{(m)}$, $i \in \{T, J\}$ given in Eqs.~\eqref{eq:payoffT} and \eqref{eq:payoffJ}.
In the remainder of this section, we will omit superscript $(m)$ for the sake of a lighter notation. 

The expected payoffs $\mathbb{E}\Big[u_{i} \Big| N_T,\,N_J\Big]$,  $i\in \{T,J\}$ can be calculated from the quantity $\mathbb{E}\Big[\chi_{i} \Big| N_T,\,N_J\Big]$, $i\in \{T,J\}$, which represents the expected outcome of the subgame (as introduced in Sec.~\ref{sec:the_packet_transmission_subgame}, $\chi_T$ and  $\chi_J=1-\chi_T$ are indicator terms for the transmission and jamming success, respectively).
We introduce quantities $N_{\rm C}\leq N_T$ and $N_{\rm B}\leq N_T$ to indicate the number of packets that node $T$ sends  over a clear and blocked (i.e., jammed) channel, respectively. Obviously, $N_T = N_{\rm C} + N_{\rm B}$, so we can easily obtain the value of $N_{\rm C}$ once we know $N_{\rm B}$. We also know that $N_C\geq 1$, as the first packet can never be jammed.
Using the law of total probability, for node $T$ we have:
\begin{equation}
    \mathbb{E}\Big[\chi_{T} \Bigm| N_{\rm T},\,N_{\rm J}\Big] = \sum_{N_{\rm C}=1}^{N_{\rm T}} \mathbb{E}\Big[\chi_{T} \Bigm| N_{\rm B}, N_{\rm T}\Big]  \prob\Big(N_{\rm B}\Bigm| N_{\rm T},\,N_{\rm J}\Big)
    \label{eq:chi_expectation}
\end{equation}
%
The first term inside the summation is the expectation of a subgame success, given the number of packets  successfully delivered and jammed during that subgame, and can be expressed as:
\begin{equation} \label{eq:chi2}
\begin{aligned}
\mathbb{E}\Big[\chi_T \!\Bigm|\! N_{\rm B}, N_{\rm T}\Big] =  \sum_{D=K}^{N_T} \sum_{d_{\rm B}=0}^D \binom{N_{\rm T}-N_{\rm B}}{D-d_{\rm B}} p_{e_{\rm C}}^{(N_{\rm T}-N_{\rm B})-(D-d_{\rm B})}&\\ \times (1-p_{e_{\rm C}})^{D-d_{\rm B}}
\binom{N_{\rm B}}{d_{\rm B}}\:p_{e_{\rm B}}^{N_{\rm B}-d_{\rm B}}\:(1-p_{e_{\rm B}})^{d_{\rm B}}\,. &
\end{aligned}
\end{equation}
%
The external summation  iterates on all possible values of the number of delivered packets $D\leq N_T$ resulting in a success.
Eq.~\eqref{eq:chi2} then splits $D$ between packets that are delivered over a jammed channel, i.e.,  $d_{\rm B}\leq D$, and those which are delivered over a clear channel, i.e., $D-d_{\rm B}$. For the two cases, the packet error probability is equal to $p_{e_{\rm C}}$ and $p_{e_{\rm B}}$, respectively, and is a function of the \gls{snr} or \gls{sinr}. 

We consider a realistic modulation, such as \gls{css}, which is used in several real underwater acoustic modems. If a different modulation is used, the only required change is in~\eqref{eq:p_bit_CSS}, while the rest of the model remains the same. The packet error probability in the case of a jammed signal, $p_{eB}$, can be computed as presented in \cite{jammerdBER_CSS}, which computes the \gls{ber} considering \gls{dqpsk} modulation in a radio frequency channel. We now adapt the definition of the \gls{ber} to the acoustic underwater scenario. The \gls{ber} for a \gls{css} signal, $p^{\text{CSS}}_{\text{bit}}$, is computed as:
\begin{equation}
    p^{\text{CSS}}_{\text{bit}} = Q(a,b)-\frac{1}{2}e^{-(a^2+b^2)/2}I_0(ab)\,
    \label{eq:p_bit_CSS}
\end{equation}
where $Q$ is the Marcum Q function, $I_0$ is the modified Bessel function of order 0, and $a$ and $b$ are defined as:
\begin{equation}
  \begin{split}
    a = \sqrt{\frac{2 E_b/N_0}{1+J_0/N_0}(1-\sqrt{0.5})}\, \\
    b = \sqrt{\frac{2 E_b/N_0}{1+J_0/N_0}(1+\sqrt{0.5})}
  \end{split}\label{eq:param_p_bit}\, ,
\end{equation}
where $E_b$ is the received energy per bit of the transmitter, $N_0$ is the noise power spectral density, and $J_0$ is the power spectral density of the jammer, given by: 
\begin{equation}
    E_{b} = \frac{\tau}{L} P_{{\rm tx},T}\,g_{\rm T}, \qquad  J_0= \frac{P_{{\rm tx},J}\,g_J}{B} \,.
    \label{eq:SNR}
\end{equation}

The \gls{sinr} is then given by $\text{SINR} = \frac{E_{b}L/\tau}{N_0 B + J_0 B} $, where $L$ is the packet length in bits, $P_{{\rm tx}, i}$ represents the transmit power of node $i\in\{T,J\}$, $B$ is the transmission bandwidth, and $g_T$ and $g_J$ model the gain of the underwater acoustic channel between $T$ and $R$ and between $J$ and $R$, respectively.
Their values  depend on the distances $d_{\rm TR}$ and $d_{\rm JR}$ to the receiver, respectively, as well as on the carrier frequency of the signal. Both noise and channel gain for an underwater acoustic channel can be computed as described in~\cite[Sec. II]{Stojanovic}.

Finally, the packet error probability $p_{e_{\rm B}}$ is given by:
\begin{equation}
    p_{e_{\rm B}} = 1-\big(1-p^{\text{CSS}}_{\text{bit}}\big)^{L},
\label{eq:uncoded_per}
\end{equation}
We also consider the packet error probability if a \gls{rs} channel code is employed~\cite{goldsmithWireless}. We analyzed the performance with an RS(127,78) with $q=7$ bits per symbol and an error correction capability of $t=24$ symbols. In this scenario, the packet is lost if more than $t$ symbols are not received correctly:
\begin{equation}
    p_{e_{\rm B}} = \sum_{i=t+1}^{N} \binom{N}{i} p_s^i(1-p_s)^{(N-i)} \,
    \label{eq:coded_per}
\end{equation}
where $p_s = 1 - \left(1 - p^{\text{CSS}}_{\text{bit}}\right)^q$ is the symbol error probability.

Finally, the second term in Eq.~\eqref{eq:chi_expectation}  can be expressed as:
\begin{equation}
\prob\Big(N_{\rm B}\Bigm| N_T,\,N_J\Big) = \frac{\binom{N_T-1}{N_{\rm B}}\binom{(2K-1)-(N_T-1)}{N_J-N_{\rm B}}}{\binom{2K-1}{N_J}}\,,\label{eq:P}
\end{equation}
where we have imposed the condition that the first transmitted packet cannot be jammed due to the signal propagation characteristics of the underwater scenario, as described in Sec.~\ref{sec:model}.
In Eq.~\eqref{eq:P} , we assume that both the transmitter and the jammer choose the slots to transmit (or jam) according to a uniform distribution among all possible $N_T$-tuples (or $N_J$-tuples) of slots. This is the choice that maximizes (for the transmitter) or minimizes (for the jammer) the probability that at least $K$ slots in the transmission are free from collision. 
This strategy pair is the \gls{ne} for the anti-coordination slot selection game we mentioned in Sec.~\ref{sec:the_packet_transmission_subgame}: since all slots after the first have the same success probability, the optimal strategy for both players is to randomly choose $N_T-1$ and $N_J$ among them. Any other strategy would be strictly dominated, since it would provide the opponent with a pattern to exploit: if $T$ chooses a slot with high probability, $J$ will try to mirror it and jam the communication more effectively. The only exception to this is the first slot, which the jammer cannot jam; it is trivial to show that a strategy that includes it with probability 1 and selects the others with uniform probability strictly dominates any others for the transmitter.

Substituting \eqref{eq:chi2} and \eqref{eq:P} into \eqref{eq:chi_expectation}, we can finally obtain the expected value of the indicator function $\chi_i^{(m)}$ and then the expected value of the payoffs $u_i^{(m)}$.

\subsection{Dynamic Programming Solution}
\label{sec:dynamicProgrammingSolution}
\begin{figure}[t!]
\centering
\hspace{-1.5cm}
\resizebox{0.40\textwidth}{!}{
\begin{tikzpicture} [>=angle 45]
 \node[draw, minimum height=7cm, minimum width=8cm, line width=0.45mm](lgd) at (-8,7) {};
 \node[draw, circle,minimum size=2.7cm, line width=0.45mm, label=right:\Huge Empty state](lgdEmpty) at (-10,8.7) {{\huge $\varepsilon$}};
 \node[draw, circle,minimum size=2.7cm, line width=0.45mm, label=right:\Huge Active state](lgdActive) at (-10,5.3) {{\huge $B_T,B_J$}};
 \node[draw, circle,minimum size=2.7cm, line width=0.45mm](00) at (0,4) {{\Huge $\varepsilon$}};
 \node[draw, circle,minimum size=2.7cm, line width=0.45mm](10) at (5,8) {{\huge $K,0$}};
 \node[draw, circle,minimum size=2.7cm, line width=0.45mm](11) at (5,4) {{\huge $K,1$}};
 \node[draw, circle,minimum size=2.7cm, line width=0.45mm](12) at (5,0) {{\huge $K,2$}};
 \node[draw, circle,minimum size=2.7cm, line width=0.45mm](20) at (10,8) {{\huge $K\!+\!1,0$}};
 \node[draw, circle,minimum size=2.7cm, line width=0.45mm](21) at (10,4) {{\huge $K\!+\!1,1$}}; 
 \node[draw, circle,minimum size=2.7cm, line width=0.45mm](22) at (10,0) {{\huge $K\!+\!1,2$}};

 \draw [->] (10) -- (00);
 \draw [->] (20) edge [bend right=75] node[swap] {} (00);
 \draw [->] (20) -- (10);
 \draw [->] (21) -- (11);
 \draw [->] (22) -- (12);
 \draw [->, dashed] (13,0) -- (11.35,0);
 \draw [->, dashed] (13,4) -- (11.35,4);
 \draw [->, dashed] (13,8) -- (11.35,8);
 \draw [->] (11) -- (00);
 \draw [->] (21) -- (10); 
 \draw [->] (22) -- (11);
 \draw [->] (21) edge [bend left=30] node[swap] {}  (00);
 \draw [->] (12) -- (00);
 \draw [->] (22) -- (10);
 \draw [->] (22) edge [bend left=75] node[swap] {} (00);
 
 \draw [->] (00) edge [loop, in=205, out=155,looseness=3] (00);
 \draw [ultra thick, loosely dotted, shorten >=1mm, shorten <=1mm] (22.south) -- +(0,-0.9cm);
 \draw [ultra thick, loosely dotted, shorten >=1mm, shorten <=1mm] (12.south) -- +(0,-0.9cm);

\end{tikzpicture}
}

\vspace{-0.5cm}
\caption{State transitions for the multistage game $\mathbb{G}$.}
\vspace{-0.3cm}
\label{fig:dynamic}
\end{figure}
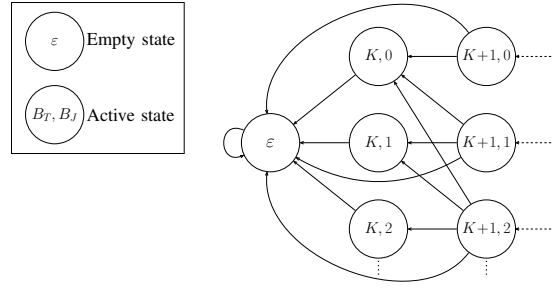

In the case of complete information, an optimal solution of the multistage game can be determined through a dynamic programming procedure.
We define the system state as $S^{(m)}\triangleq \left(B_{T}^{(m)},B_{J}^{(m)}\right)$, where $B_i^{(m)}$ is limited by the initial battery level $B_{i}^{(0)}$ of player $i\in\{T,J\}$. The state space is then defined as $\mathcal{S}=\{0,\ldots,B_{T}^{(0)}\}\times\{0,\ldots,B_{J}^{(0)}\}$.
If $\Gamma>1$, the payoff in state $S^{(m)}$ takes the payoff of the future $\Gamma-1$ subgames into account. The game ends when the transmitter's battery level is too low to transmit at least $K$ packets, i.e., when $B_{T}^{(m)}<K$. We can aggregate all states that satisfy the ending condition into a final state $\varepsilon$ and define its payoff as:
\begin{equation}\label{eq:laststep}
    U_i^{(m)}\left(\Gamma\Bigm| S^{(m)}=\varepsilon\right)=0\quad\forall\,i,\Gamma\,.
\end{equation}
We can now compute $\mathbb{E}\left[U_i^{(m)}\left(\Gamma\right)\Bigm| S^{(m)}\right]$ recursively for all other states, considering that the battery charge can never increase, hence $B_{i}^{(m+1)}\leq B_{i}^{(m)}\,\forall i,m$. It is:
\begin{equation} \label{eq:recursive}
\begin{aligned}
    \mathbb{E}\left[U_i^{(m)}\left(\Gamma\right)\Bigm| S^{(m)}\right]=\mathbb{E}\left[u_i^{(m)}\Bigm|  S^{(m)}\right]+&\\
     \lambda \sum_{S\in\mathcal{S}} \mathbb{E}\left[U_i^{(m+1)}\left(\Gamma-1\right)\Bigm| S\right]\prob\left(S^{(m+1)}=S\Bigm| S^{(m)}\right).&
\end{aligned}
\end{equation}
The payoff in a state is thus computed as the expected payoff $\mathbb{E}[u_i^{(m)}]$ obtained in the subgame corresponding to that state plus the payoff that is expected to be obtained in the next $\Gamma-1$ subgames, with an exponential discount factor $\lambda$ (see~\eqref{eq:fullpayoff}). This latter term is calculated by averaging over each possible next state $S^{(m+1)}$ weighed by the probability of transitioning to that state.
For a given pair of strategies $(s_T,s_J)$, such state transition probability is given by:
\begin{equation} \label{eq:probstate}
\begin{aligned}
  \prob\Big( S^{(m+1)}\!=\!(B_T,B_J) \Bigm| S^{(m)}\Big) =&\\ \Phi_{s_T}\left(B_T^{(m)}-B_T^{(m+1)}\right)
    \Phi_{s_J}\left(B_J^{(m)}-B_J^{(m+1)}\right).&
\end{aligned}
\end{equation} 
By substituting~\eqref{eq:probstate} into~\eqref{eq:recursive}, we have a full recursive formulation for the expected long-term payoff $\mathbb{E}[U_i^{(m)}(\Gamma)]$ for any strategy pair. Once the payoff bimatrix is thus constructed, the Lemke-Howson algorithm can be used to find the mixed \gls{ne}~\cite{lemke1964equilibrium}. By starting from the lowest states and calculating the expected payoffs $\mathbb{E}\left[U_i^{(m)}(\gamma)\right],\,\gamma\!\in\!\{1,\!\ldots,\!\Gamma\}$, the game can be solved completely.
Fig.~\ref{fig:dynamic} shows the state transition graph for the multistage game $\mathbb{G}$.
Transitions are allowed from bottom to top and from right to left, as a consequence of nodes $T$ or $J$ consuming energy to send packets or jam slots, respectively.
The game ends at stage $h\in\mathbb{N}$ when state $\varepsilon$ is reached, i.e., $B_{T}^{(h)}<K \leq B_{T}^{(h-1)}$.
Notice that, if the battery of $J$ empties before $T$'s, the game evolves in the limit condition of $T$ playing against the channel. 

\subsection{Analytical performance evaluation}
After computing the strategies, we can evaluate the expected lifetime $\mathbb{E}\left[L|S^{(m)}\right]$ of the transmitter node, defined as the number of blocks that it can transmit (either successfully or not), i.e., the number of subgames that will be played before its battery is depleted. Using~\eqref{eq:probstate}, we define the expected lifetime for state $S^{(m)}$ recursively:
\begin{multline}\label{eq:lifetime}
\mathbb{E}\!\left[\!L\!\Bigm|\!S^{(m)}\!\right]\!=\!\sum_{B_J=0}^{B_J^{(m)}}\!\sum_{B_T=0}^{B_T^{(m)}-K}\!\left(\!1\!+\!\mathbb{E}\!\left[L\!\Bigm|\!S^{(m+1)}\!=\!(B_T,B_J)\right]\right)\\ \times 
\prob\left( S^{(m+1)}\!=\!(B_T,B_J) \Bigm| S^{(m)}\right).
\end{multline}
The lifetime takes into account the subgame $(m)$, which is summed to the expected lifetime of each possible next state $S^{(m+1)}$, weighed by its probability. Since the game ends in state $\varepsilon$, we can now define the base step of the recursive formulation:
\begin{equation}
 \mathbb{E}\left[L\Bigm|S^{(m)}=\varepsilon\right]=0.
\end{equation}

We can also derive the expected success probability $P_S\left(S^{(m)}\right)$ using the same reasoning. The success probability for the current subgame is averaged with the success probability in future states, weighed by the expected lifetime and the probability of reaching those states using~\eqref{eq:lifetime}:
\begin{equation}
\begin{aligned}
P_S\left(S^{(m)}\right)=\sum_{N_T=K}^{2K}\sum_{N_J=0}^{2K-1}\prob\left( S^{(m+1)}=S_{N_T,N_J}^{(m+1)} \Bigm| S^{(m)}\right)&\\ \times
\frac{\mathbb{E}\left[\chi_T|N_T,N_J\right]+\mathbb{E}\left[L\Bigm|\left(S_{N_T,N_J}^{(m+1)}\right)\right]P_S\left(S^{(m+1)}\right)}{1+\mathbb{E}\left[L\Bigm| S^{(m+1)}_{N_T,N_J}\right]},&
\end{aligned}
\end{equation}
where $S_{N_T,N_J}^{(m+1)}=\left(B_T^{(m)}-N_T,B_J^{(m)}-N_J\right)$. The base step is the same as for the lifetime:
\begin{equation}
 P_S\left(\varepsilon\right)=0.
\end{equation}

\begin{figure*}[t]
\centering
\includegraphics[width=0.7\textwidth]{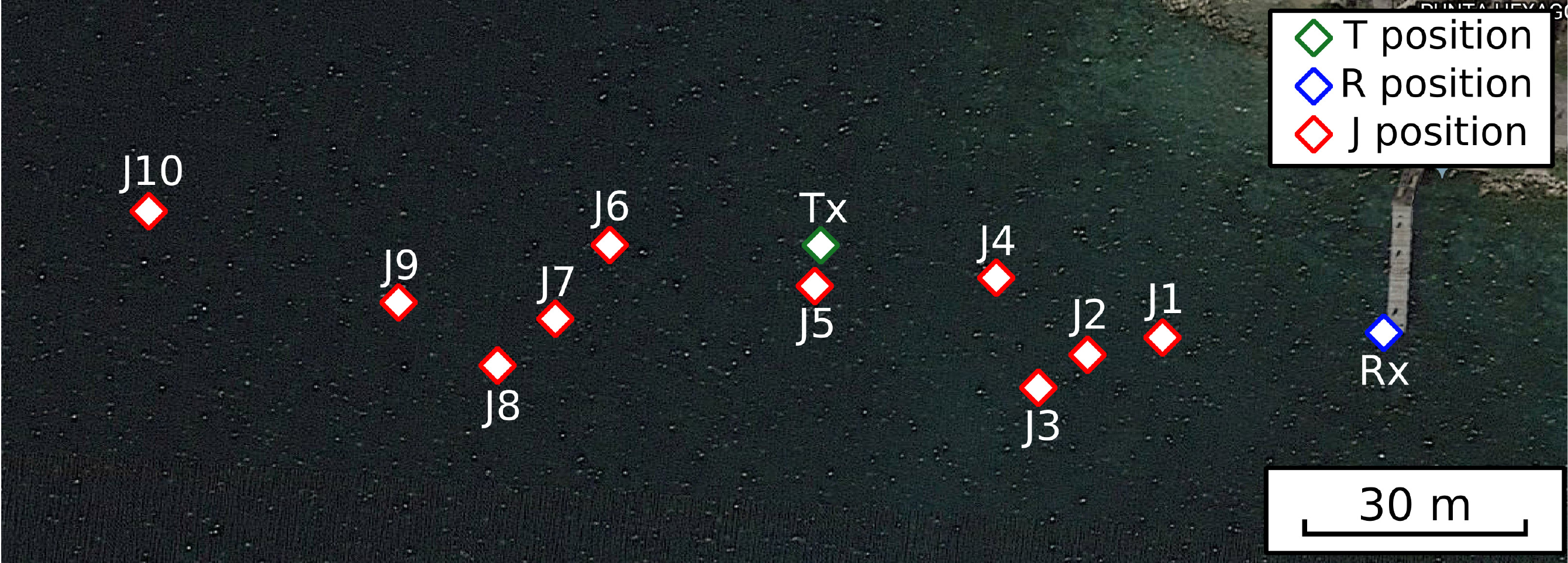}
\caption{Node deployment in the Garda lake. The figure reports all the positions (red diamond) in which the jammer node was placed during the experiment. Transmitter position (green diamond) and receiver position (blue diamond) are reported as well.} 
\label{fig:sea_trial_scenario}
\vspace{-3mm}
\end{figure*}

\subsection{Computational complexity} \label{sec:complexity}
The computation of the optimal strategies requires the dynamic programming approach described in Sec.~\ref{sec:dynamicProgrammingSolution}, starting from the states with the lowest battery level and exploiting the results to calculate the strategy for the subsequent ones. The solution of the game in a given state $S=(B_T,B_J)$ then requires the knowledge of the expected payoff for the current subgame and for future ones, which are then given as input to the Lemke-Howson algorithm to find the \gls{ne}. If we denote the complexity of finding the expected payoff of a subgame as $M_{\text{sub}}$, and the complexity of the Lemke-Howson algorithm as $M_{\text{LH}}$, the overall complexity of the solution in state $S$ is $O(B_T B_J M_{\text{sub}}M_{\text{LH}})$.

While the Lemke-Howson algorithm is efficient in practice, its worst-case complexity has been shown to be $M_{\text{LH}}\sim O(2^{3K})$~\cite{savani2006hard}, as it depends directly on the length of the longest pivoting path in the strategy space. \footnote{We remind the reader that $K$ is the number of information packets in a burst.}

We now compute $M_{\text{sub}}$, the last term of the overall complexity formula.
In order to find the expected payoff in a subgame, the nodes need to solve~\eqref{eq:chi_expectation} for each pair of possible moves $(N_T,N_J)$. The solution of~\eqref{eq:chi_expectation} requires to compute $\mathbb{E}\left[\chi_T\!\mid\!N_{\rm C}\right]$ using~\eqref{eq:chi2} $N_T$ times, and~\eqref{eq:chi2} requires $O(K^2)$ operations in the worst case. Therefore, the complexity of~\eqref{eq:chi_expectation} is $O(K^3)$, and computing the expected payoff for all the $(K+1)(2K-1)$ possible pairs of moves is $O(K^5)$.

The overall time to find the solution of the game in state $S$ is then $O(B_T B_J K^5 2^{3K})$, which is clearly intractable for a computationally limited underwater node. However, the optimal strategies can be computed offline and loaded in the agent as a simple lookup table, reducing the complexity to a simple memory read.

\section{Scenario settings}\label{sec:scenario}
We evaluate the performance of the optimal strategies by studying the energy consumption and the \gls{pdr} of $T$ in two scenarios, a model-based one and an experimental one. We set up the two scenarios using the same transmitter and scenario parameters, trying to make them as comparable as possible: for this reason, the relative positions of the three nodes (transmitter, jammer, and receiver) are the same in both scenarios.
We considered a carrier frequency equal to $26$~kHz and a bandwidth of $16$~kHz. The transmit power was the same for both transmitter and jammer, namely $P_{{\rm tx}, i} =$~180~dB~re~1$\mu$Pa, $i\in\{T,J\}$. However, the two scenarios used different packet error probabilities, derived from a theoretical model and a lake experiment, respectively.

\subsection{Model-based scenario} \label{sec::scenario.model}
In this scenario, jammer and transmitter are trained and evaluated using the uncoded \gls{css} modulation with \gls{dqpsk}; the packet error probability is given in~\eqref{eq:uncoded_per}. As mentioned above, the considered propagation model is described in~\cite{Stojanovic}, where only the \gls{los} component is considered.  The wind speed, shipping factor and geometrical spreading factor are set to 3 m/s, 1, and 1.75, respectively.  We remark that the settings are different from our previous work: we used a different modulation, which leads to a different packet error probability formula. However, the game theoretic model works independently of these settings. The channel settings (with few reflections and a strong line of sight component) are optimistic, as real scenarios in shallow water often have strong reflections and environmental noise. The parameters of the model are summarized in Table~\ref{tab:parameters}.

\subsection{Experimental settings}\label{ssec:sea}
\begin{figure}[t]
\centering
\includegraphics[width=0.85\columnwidth]{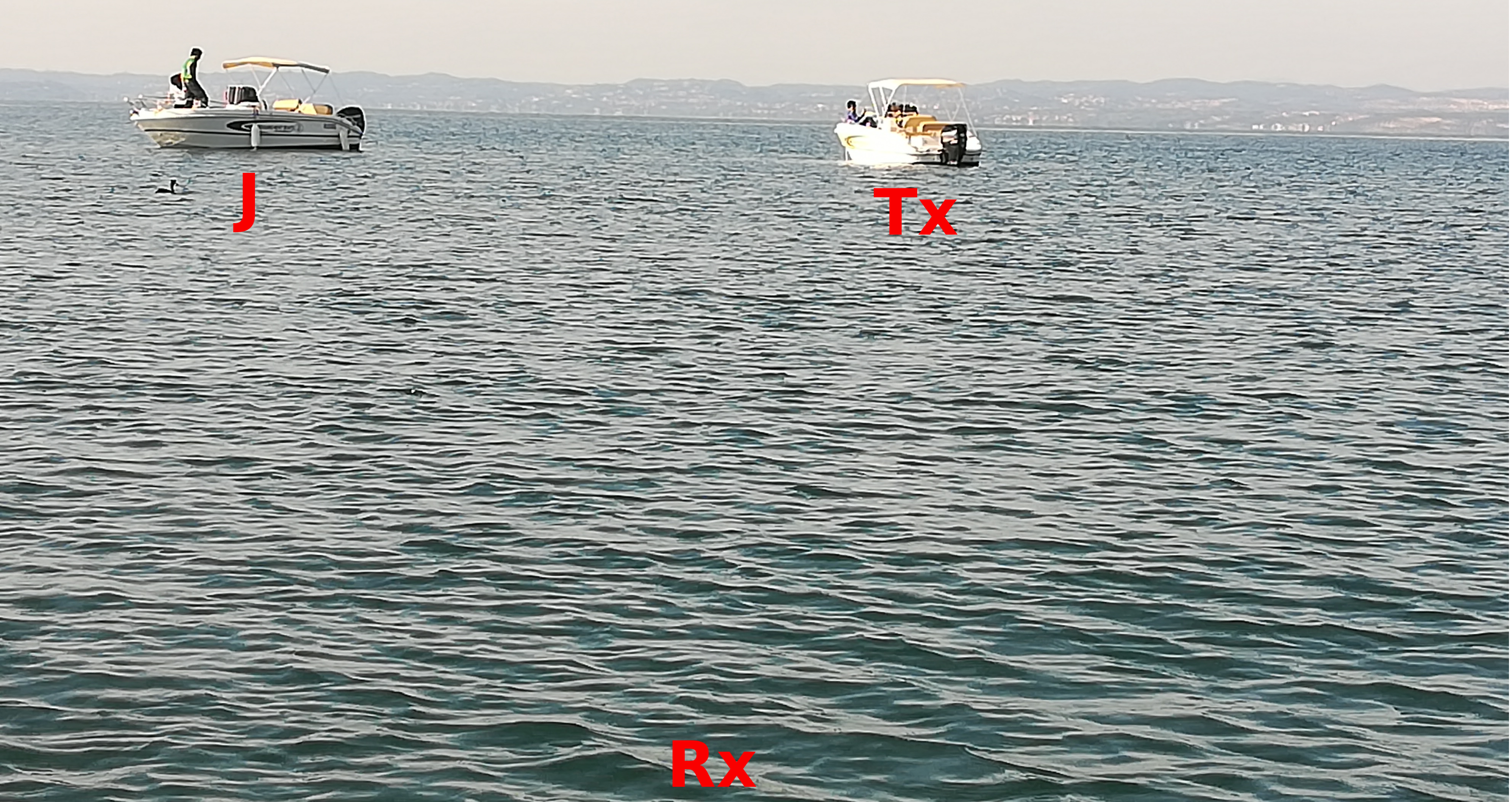}
\caption{Picture of the experiment taken from the receiver node station when the jammer was in position J5 (Figure~\ref{fig:sea_trial_scenario}).} \label{fig:sea_trial}
\end{figure}

\begin{figure}[t]
\centering
\includegraphics[width=0.7\columnwidth]{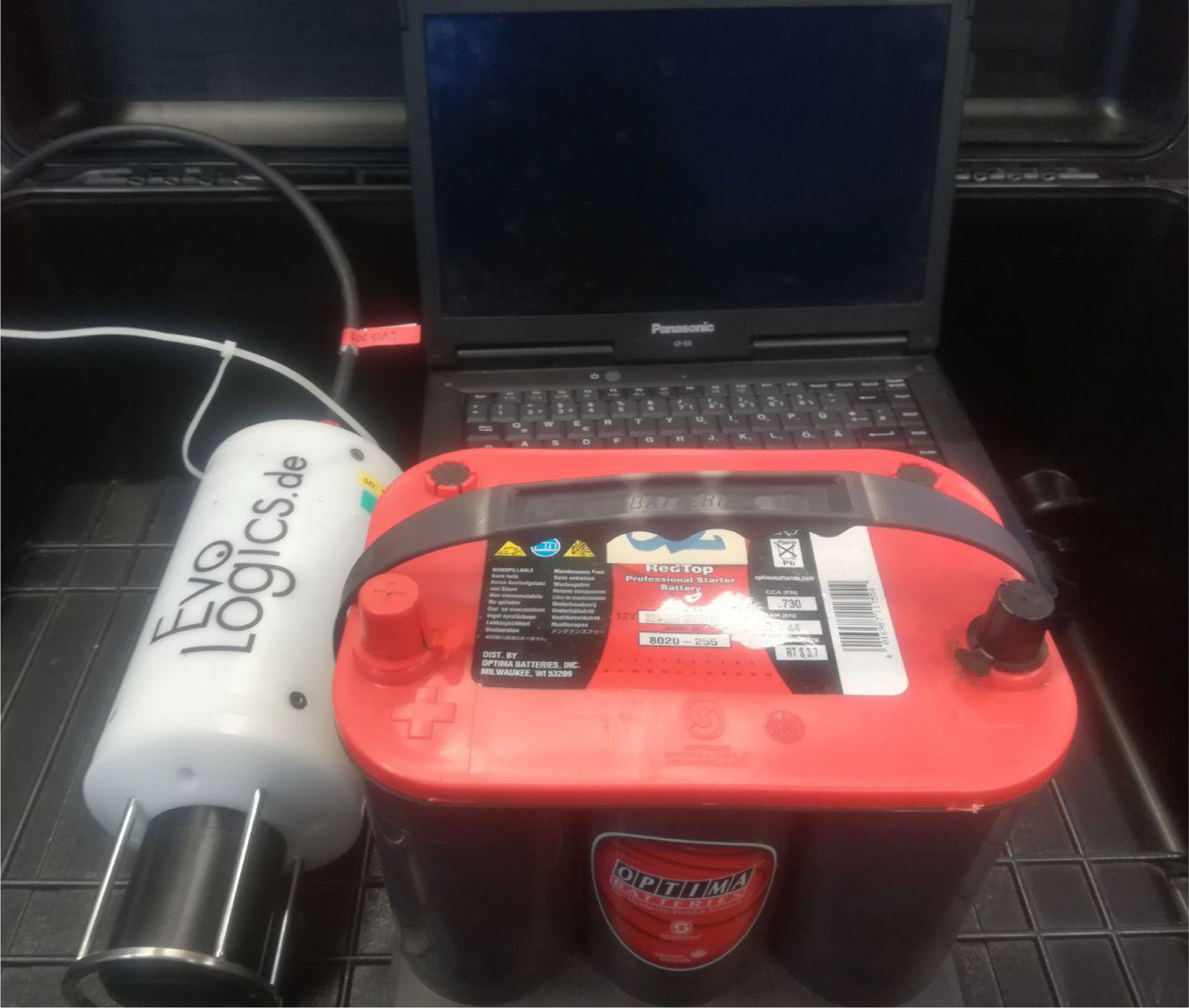}
\caption{Picture of the apparatus used in the experiment. Each node was equipped with batteries, a laptop and an acoustic modem.} \label{fig:sea_trial_eqipment}
\end{figure}

The lake experiment took place in the Garda lake on Thursday 17th October 2019, just off the Bardolino town coastline. 
The weather was sunny, and the maximum wind speed we experienced during the experiment was 8 m/s. Most of the waves were caused by the motion of the surrounding ships: shipping activity was very heavy, as our network was deployed at only 500 m from the Bardolino ferry station, and the receiver node was placed close to a boat rental service. All the measurements were performed from 10 AM to 4 PM. 
The experimental setup was composed of 3 nodes equipped with EvoLogics S2C R 18/34 WiSE modems~\cite{S2CR1834WISE}: the receiver was deployed from a floating pier (N 45.549108, E 10.715181), the transmitter from a working boat anchored 80 meters west of the receiver (N 45.549165, E 10.714172), and the jammer from a working boat placed at different locations, between 20 and 180 meters west of the receiver. The map of the node positions is shown in Fig.~\ref{fig:sea_trial_scenario}, Fig.~\ref{fig:sea_trial} is a photo of the scenario from the receiver's perspective, and Fig.~\ref{fig:sea_trial_eqipment} shows the equipment used for each node in the experiment. The water depth was 4~m at the receiver, 10~m at the transmitter, and varied from 4 to 15 meters at the jammer, depending on its location. All nodes were deployed at a depth of 2~m, and both $J$ and $T$ were sending signals with an acoustic power of 180~dB~re~1$\mu$Pa. 
Both modems deployed from the Tx and the Rx stations used the standard EvoLogics firmware, while node $J$ was transmitting continuous signals at 1~kbps by using the low-level EvoLogics firmware, described in~\cite{lowlevel_wuwnet17}. Every 2 seconds, $T$ sent one instant message packet with a payload length of 64 Bytes at the same bitrate of $J$. Together with the EvoLogics header and coding used by the standard EvoLogics, the packet duration was approximately 0.86~s (value provided by the modem at the moment of the reception). In order to prevent $T$'s transmissions from being blocked by the reception of $J$'s signals (as the acoustic modems are, for their nature, half-duplex devices), $T$ was set in deaf mode, i.e., its receiver unit was disabled. Both $T$ and $J$ used a \gls{qpsk} modulation, with each symbol spread to the whole bandwidth (using the so-called sweep-spread carrier (S2C) technology).
%

\begin{table}[t!]
	\caption{Parameters setting.}
	\vspace{-0.3cm}
	\footnotesize
	\centering
	\begin{tabular}{ll}
		\Xhline{2\arrayrulewidth}
		Parameter & Value \\ \Xhline{2\arrayrulewidth}
		Modem carrier freq & 26~kHz \\
		Modem bandwidth & 16~kHz \\
		Modem bitrate & 1~kbps \\
		Payload length & 64~Bytes \\
		T and J $P_{\rm tx}$ & 180~dB~re~1$\mu$Pa \\
		$p_{e_{\rm C}}$ (lake exp) & 0.04 \\
		$p_{e_{\rm C}}$ (models) & 0 \\
		Spreading factor k & 1.75 \\
		Shipping factor s & 1 \\
		Wind speed w & 3 m/s \\
		\Xhline{2\arrayrulewidth}
	\end{tabular}
	\label{tab:parameters}
	\vspace*{-0.4cm}
\end{table}
\section{Numerical evaluation}\label{sec:results}
In this section we report and assess the results for both the model-based and the lake experiment scenarios described in Section \ref{sec:scenario}.

\subsection{Model-based scenario results}\label{sec:sim_results}

\begin{figure}[t]
 \centering
 \setlength\fwidth{0.88\columnwidth}
 \setlength\fheight{0.43\columnwidth}
 \input{./figures/regions.tex}
 \vspace{-0.4cm}
\caption{Blocked channel packet error rate $p_{e_{\rm B}}$ for a jammed slot as a function of the distance $d_{\rm JR}$ between J and R when the distance between $T$ and $R$ is  $d_{\rm TR}=78$~m, using the uncoded model.} 
\label{fig:PER}
\vspace{-2mm}
\end{figure}
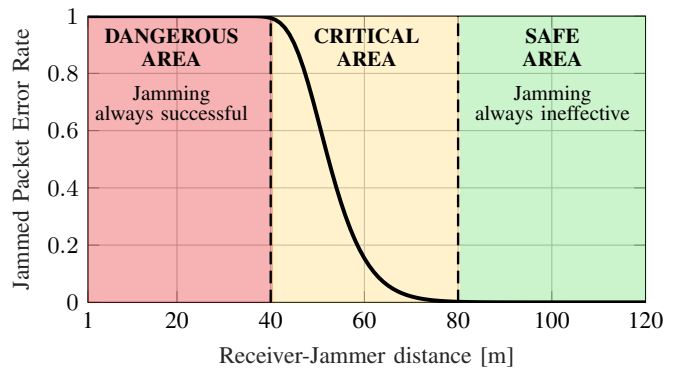

Based on the position of the jammer, we can distinguish three regions in the underwater area, as shown in Fig.~\ref{fig:PER}. When the jammer is close to the receiver, any jammed packet is almost surely lost, as the received jamming signal is powerful enough to cause errors in the transmission. In our system scenario, this situation happens when the receiver-jammer distance is less than 40~m. Conversely, when the jammer is far from the receiver, its attack is completely ineffective, as the legitimate signal is much stronger; in our case, this happens when $J$ is farther than 80~m from $R$. Between these two extremes, an appropriate strategy might significantly improve the performance: it is interesting to investigate how the game evolves in the critical region (where $d_{\rm JR}\!\in\![40,80]$~m in our scenario), and which distances yield a successful game for $T$.

Although this performance figure refers to a specific combination of transmission power and modulation, a different configuration would still lead to the definition of the three regions, but at different distances between the transmitter and the jammer~\cite{jamming_uw}.

\begin{figure}[t]
 \centering
 \setlength\fwidth{0.88\columnwidth}
 \setlength\fheight{0.43\columnwidth}
 \input{./figures/Psucc_dist_ws.tex}
\vspace{-7mm}
\caption{Success probability in a single subgame as a function of $d_{\rm JR}$ using the uncoded model, for different values of $\Gamma$ when $\alpha=0.4$.} \vspace{-0.2cm}
\label{fig:Psucc_ws}
\vspace{-2mm}
\end{figure}
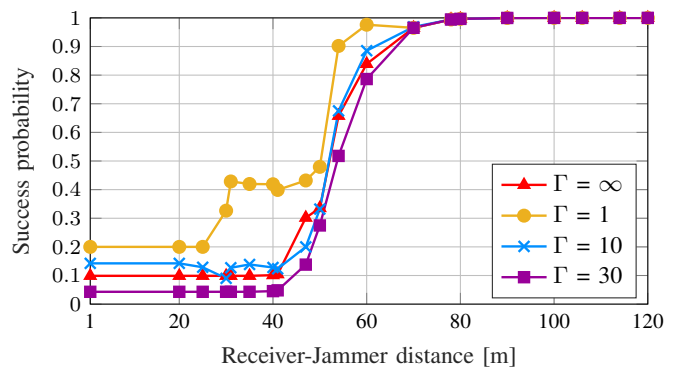

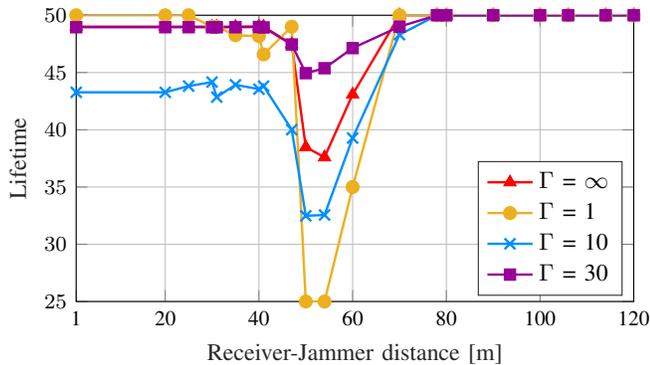
\begin{figure}[t]
 \centering
 \setlength\fwidth{0.88\columnwidth}
 \setlength\fheight{0.43\columnwidth}
 \input{./figures/lifetime_distance_ws.tex}
\vspace{-4mm}
\caption{Transmitter's lifetime as a function of $d_{\rm JR}$ using the uncoded model, for different values of the time horizon $\Gamma$ when $\alpha=0.4$.} 
\vspace{-0.2cm}
\label{fig:Lifetime_ws}
\end{figure}

This partition is also clear from Fig.~\ref{fig:Psucc_ws}, which shows the transmission success probability of a subgame as a function of the distance between $J$ and $R$. The success probability is close to 1 when the jammer is far away, and quickly drops when it gets closer than 80 m. It is interesting to note that the success probability when the jamming node is close decreases for longer time horizons; in this case, $T$ tries to save energy while still transmitting, and a shorter window leads to a more aggressive policy.
However, agents with a longer time horizon can avoid suboptimal choices. The jammer is particularly affected by this short-sightedness, as its reward function does not explicitly have a penalty for energy expenditure, and it will waste energy if its horizon is too short, quickly exhausting its own battery. This causes a temporary drop in the success probability, which is quickly reversed when the jammer depletes its battery and tries to fight a lost battle against the transmitter. In fact, a short time horizon corresponds to both a higher success probability and a higher lifetime for the transmitter, as shown in Fig.~\ref{fig:Lifetime_ws}. Since the initial jammer battery $B_J^{(1)}$ is set to 200 packets, $\Gamma=30$ is the only value that ensures that the jammer will not act in a myopic way. We remark that in this case, simply switching to a short-term strategy will not benefit the legitimate transmitter: since the long-term result is the \gls{ne}, choosing any other strategy will decrease its expected payoff even further. It is interesting to note that the infinite horizon jammer also suffers from this issue, since its temporal discount $\lambda=0.9$ is small enough to make it weigh present rewards more than heavy future losses. For the rest of this analysis, we will consider the scenario in which $\Gamma=30$.

\begin{figure}[t]
 \centering
 \setlength\fwidth{0.88\columnwidth}
 \setlength\fheight{0.43\columnwidth}
 \input{./figures/Psucc_dist_alpha.tex}
\vspace{-4mm}
\caption{Success probability in a single subgame as a function of $d_{\rm JR}$ using the uncoded model, for different values of $\alpha$ when $\Gamma=30$.} \vspace{-0.2cm}
\label{fig:Psucc_alpha}
\vspace{-0.2cm}
\end{figure}
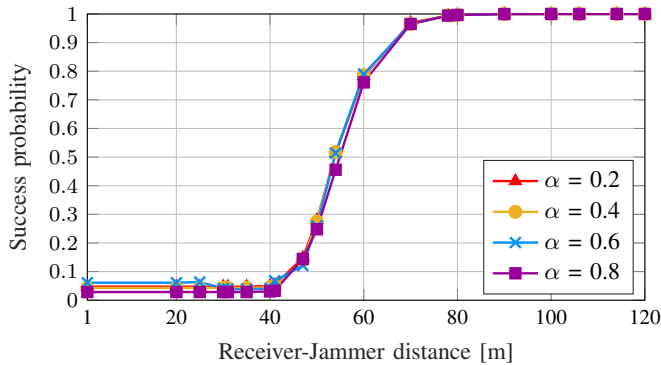

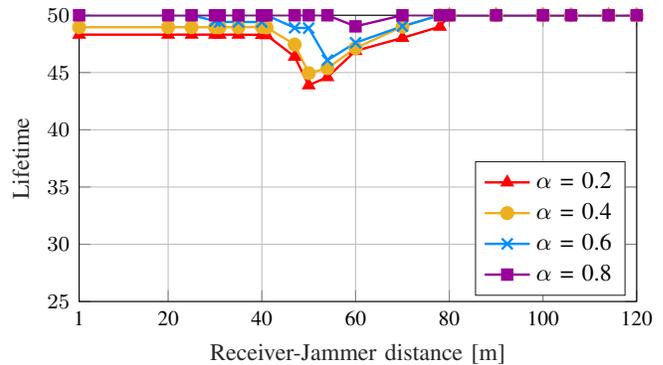
\begin{figure}[t]
 \centering
 \setlength\fwidth{0.88\columnwidth}
 \setlength\fheight{0.43\columnwidth}
 \input{./figures/lifetime_distance_alpha.tex}
\vspace{-4mm}
\caption{Transmitter's lifetime as a function of $d_{\rm JR}$ using the uncoded model, for different values of $\alpha$ when $\Gamma=30$.} 
\label{fig:Lifetime_alpha}
\vspace{-0.2cm}
\end{figure}

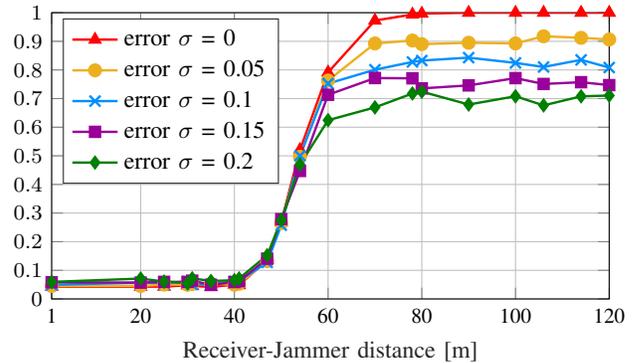
\begin{figure}[t]
	\centering
	\setlength\fwidth{0.88\columnwidth}
	\setlength\fheight{0.43\columnwidth}
	\input{./figures/Psucc_sensitivity.tex}
	\vspace{-4mm}
	\caption{Success probabilities for different values of the error standard deviation $\sigma$ as a function of $d_{\rm JR}$ using the uncoded model, for $\alpha=0.4$ and $\Gamma=30$.} 
	\label{fig:Psucc_sensitivity}
	\vspace{-0.2cm}
\end{figure}

The aggressiveness of a long-sighted jammer seems to have little effect on the results: as Fig.~\ref{fig:Psucc_alpha} shows, lower values of the parameter $\alpha$ correspond to a slightly higher success probability, but the curves are very close. A jammer close to the receiver can reduce the transmission success probability to less than 10\%, but the aggressiveness parameter only has a significant impact on the success probability in the critical region. Since $K=4$ and $B_{T,0}=200$, the maximum lifetime of $T$ is 50 subgames, and is reached when $T$ does not add any \gls{fec}. The minimum lifetime is 25 subgames, in the case in which  $T$ always sends $2K$ packets, providing the maximum possible protection to its payload. Fig.~\ref{fig:Lifetime_alpha} confirms that there is a downside to aggressiveness: more conservative nodes with a higher $\alpha$ have a slightly longer lifetime in the critical region. Naturally, the lifetime is maximized when $d_{\rm JR} > 80$~m, i.e., when the jammer no longer affects the packet reception. This result holds for each value of $\alpha$; in this situation, since almost all packets are received correctly, the best strategy for the transmitter is to send exactly $K$ packets, in order to minimize the energy consumption. Naturally, the critical area definition depends on the transmission power and modulation, and its boundaries can be different in other scenarios.

We also note that the lifetime decreases when the jammer is in the critical region, where strategies have a significant impact on the outcome of the game, and transmitters have to behave more aggressively to maximize their payoff. Accordingly, the decrease is far less pronounced for higher values of $\alpha$ and longer time horizons.

We also perform a sensitivity analysis by running a Monte Carlo simulation of this scenario, changing the error probabilities $p_{e_{\rm C}}$ and $p_{e_{\rm B}}$ randomly at each run. We set a threshold for the blocked channel error probability, so that it is never lower than the clear channel error probability, and add two independent Gaussian components with zero mean and standard deviation $\sigma$ to each component. In this case, the choices of the two players become suboptimal, since they are operating with an incorrect model of the environment. Node lifetime is not affected, since the nodes make the same choices, but the success probability is, as Fig.~\ref{fig:Psucc_sensitivity} shows. The effect is interesting, and most noticeable outside the critical region: when the jammer is very close to the receiver, the success probability slightly improves as $\sigma$ grows, while the opposite happens (with much larger effects) when the jammer is far. This might be due to the threshold effect, as the packet error probability cannot be lower than 0 or higher than 1: in this case, the errors are biased. In the critical region, the model error has a slightly negative effect on the success probability, favoring the jammer. 

\subsection{Experimental scenario results} \label{ssec:trial_results}

\begin{figure}[t]
 \centering
 \setlength\fwidth{0.88\columnwidth}
 \setlength\fheight{0.43\columnwidth}
 \input{./figures/PER_seaTrial.tex}
\vspace{-2mm}
\caption{Blocked channel packet error rate $p_{e_{\rm B}}$ for different channel models as a function of $d_{\rm JR}$, for $\alpha=0.4$ and $\Gamma=30$.} 
\label{fig:per_sea}
\vspace{-0.2cm}
\end{figure}
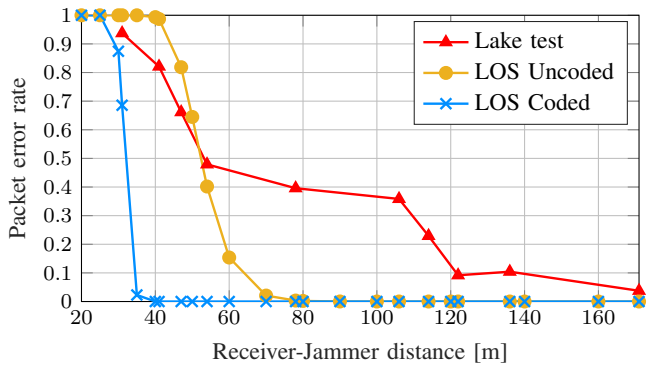

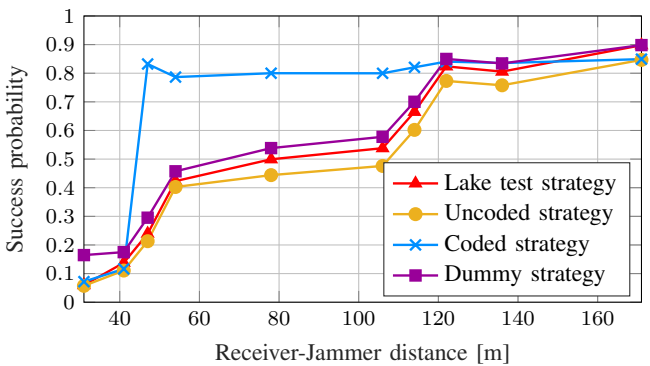
\begin{figure}[t]
 \centering
 \setlength\fwidth{0.88\columnwidth}
 \setlength\fheight{0.43\columnwidth}
 \input{./figures/Psucc_dist_seaTrial.tex}
\vspace{-2mm}
\caption{Success probability for different strategies as a function of $d_{\rm JR}$ in the lake scenario, for $\alpha=0.4$ and $\Gamma=30$.} 
\label{fig:successp_sea}
\vspace{-0.2cm}
\end{figure}

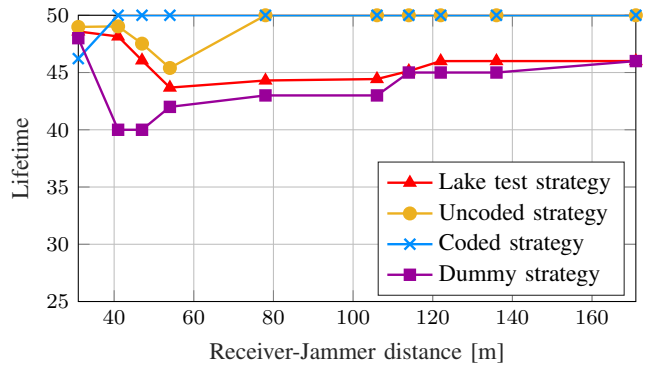
\begin{figure}[t]
 \centering
 \setlength\fwidth{0.88\columnwidth}
 \setlength\fheight{0.43\columnwidth}
 \input{./figures/lifetime_distance_seaTrial.tex}
\vspace{-2mm}
\caption{Transmitter's lifetime for different strategies as a function of $d_{\rm JR}$ in the lake scenario, for $\alpha=0.4$ and $\Gamma=30$.} 
\label{fig:Lifetime_sea}
\vspace{-0.2cm}
\end{figure}

Fig.~\ref{fig:per_sea} shows the packet error rate measured at different distances in the lake experiment, and compares it to those obtained with the the coded and uncoded \gls{los} channel models. The three curves have a sigmoid-like shape, but the real results have a relatively high packet error rate even when the jammer is far from the receiver. The uncoded packet error rate curve is similar to the measured curve for $d_{\rm JR}\leq60$ m, while the coded packet error curve is completely different, as the jammer is already supposed to be completely ineffective at a distance $d_{\rm JR}\simeq40$ m. This shows that the channel model we used in the theoretical analysis was extremely optimistic, with a strong \gls{los} component and a very low ambient noise. The real propagation environment is instead much more hostile, and as a result the packet error probability is generally higher (even though the communication system used channel coding), especially in a shallow water scenario akin to the one experienced during the lake experiment.

In this scenario, we consider the lake experiment curve as the real packet error rate, and test players that devise their strategies according to different internal models: since it would be impractical to perform sea trial scenarios before deployment, nodes may have to be trained based on a theoretical model, but the difference between the models and reality may have effects on the performance, which we will analyze in the following. We also consider a dummy jammer which always jams $K+1$ slots, allowing the transmitter to find the best response: since the dummy strategy is not necessarily a best response, it is worse for the jammer than the \gls{ne} solution. This case is used as a baseline, since it is the most favorable for the transmitter.

Fig.~\ref{fig:successp_sea} shows how using a very optimistic model of the packet error probability leads to an unbalanced scenario: the jammer, convinced that its actions will have little or no effect, saves energy by limiting its transmissions. 
Most of the time the transmitter has a free channel and just has to contend with the ambient noise.
The players using the uncoded model, which is much closer to the real packet error probability curve, reach a similar equilibrium. Finally, the dummy jammer strategy is actually close to optimal at long distances, while it allows much more data to get through when the jammer is close.

In general, the transmitter almost always chooses a conservative strategy, as Fig.~\ref{fig:Lifetime_sea} shows. Since the correct packet error probabilities are higher than those predicted by the models, and particularly the coded one, the lifetime of the node with the correct model decreases as it transmits slightly more redundancy. This is also true for the dummy jammer case, as its strategy of jamming $K+1$ slots in each subgame is quite aggressive.

\section{Conclusions and future work}\label{sec:conclusions}

In this work, we modeled and analyzed an underwater jamming attack aimed at disrupting the victim's communication, as well as depleting its battery. The legitimate transmitter can leverage packet-level coding to protect its transmissions from the jammer, at the cost of an additional energy expenditure. We model the attack using game theory and derive the optimal strategies in various scenarios, assuming that the jammer and the legitimate transmitter are two rational players with complete knowledge about the adversary, playing a zero-sum game. The simulation results highlight three regions where the jamming attack is almost always successful, depends on the strategies of the two players, or is ineffective, respectively. The critical region in which the strategies can make a difference is the one in which the uncertainty over whether or not jammed packets can be received is high.

In addition, we analyze what happens when the nodes do not have complete information about the environment, or consider short-term goals. Reducing the players' horizon gives an advantage to the transmitter, as the jammer will waste energy in the early stages of the attack. Conversely, adding a random error to the channel model advantages the jammer in the ineffective region. Finally, we consider a more realistic scenario, in which the packet error probability is determined by the results of an experiment performed in Lake Garda. We use this realistic channel to analyze what happens when the whole model of the channel is wrong: in one case, a jammer using the wrong model performs worse than a simple dummy strategy that always tries to jam the same number of packets.

Although the analytical solution is based on the simplifying assumption of complete information available at the two players, it still sheds light on the dynamics in this scenario. The results of the model comparison analysis show that relaxing the complete information assumption, i.e., considering a Bayesian incomplete information game, would significantly improve the applicability of the model, making the nodes' strategies more robust to errors in the initial assumptions~\cite{chiariotti2020bayesian}. Other possible avenues of future research include a wider action space, which might include other defense mechanisms such as power control or frequency hopping, and the extension of the framework to a network scenario with multiple transmitters, receivers, and jammers.



\section*{Acknowledgment}

This work was partially supported by the Bundeswehr Technical Center for Ships and Naval Weapons, Maritime Technology and Research (WTD 71), Kiel, Germany. We thank Roberto Francescon and Emanuele Coccolo for their practical support and participation in the lake experiment, and Chiara Pielli, Marco Giordani, and Nicola Laurenti for their work on the preliminary version of this paper.


\bibliographystyle{IEEEtran}
\bibliography{bibliography_final}

\end{document}

%% file: figures/regions.tex
%
%
\definecolor{orange}{rgb}{1,0.75,0}%
\definecolor{orange_D}{rgb}{1.0000,0.5,0}%
\definecolor{cyan}{rgb}{0,0.6,0.64}%
\definecolor{red}{rgb}{0.9,0,0}%
\definecolor{green}{rgb}{0,0.8,0}%
\begin{tikzpicture}
\pgfplotsset{every tick label/.append style={font=\small}}
\tikzstyle{dotted} = [dash pattern=on \pgflinewidth off 0.5mm] 
\tikzstyle{dashed} = [dash pattern=on 7.5*0.8*0.8pt off 7.5*0.4*0.8pt]
\tikzstyle{dashed2} = [dash pattern=on 7.5*0.5*0.8pt off 7.5*0.25*0.8pt]
\tikzstyle{dashdotted} = [dash pattern=on 7.5*0.8*0.6pt off 7.5*0.8*0.3pt on \the\pgflinewidth off 7.5*0.8*0.3pt]
\tikzstyle{dotted2} = [dash pattern=on 7.5*0.8*0.2pt off 7.5*0.8*0.1pt]

\begin{axis}[%
width=0.951\fwidth,
height=\fheight,
at={(0\fwidth,0\fheight)},
scale only axis,
xmin=1,
xmax=120,
xtick={1,20,40,60,80,100,120},
xticklabels={1,20,40,60,80,100,120},
xlabel style={font=\small\color{white!15!black}},
xlabel={Receiver-Jammer distance [m]},
ymin=0,
ymax=1,
ylabel style={font=\small\color{white!15!black}},
ylabel={Jammed Packet Error Rate},
axis background/.style={fill=white},
xmajorgrids,
ymajorgrids,
]
\begin{scope}[]
\fill[red,opacity=.3] ({rel axis cs:0,0}) rectangle ({rel axis cs:0.331,1});
\fill[orange,opacity=.2] ({rel axis cs:0.331,0}) rectangle ({rel axis cs:2/3,1});
\fill[green,opacity=.2] ({rel axis cs:2/3,0}) rectangle ({rel axis cs:1,1});
\end{scope}
\addplot [color=black, line width=1.5pt]
  table[row sep=crcr]{%
1	1\\
2	1\\
3	1\\
4	1\\
5	1\\
6	1\\
7	1\\
8	1\\
9	1\\
10	1\\
11	1\\
12	1\\
13	1\\
14	1\\
15	1\\
16	1\\
17	1\\
18	1\\
19	1\\
20	1\\
21	1\\
22	1\\
23	1\\
24	1\\
25	1\\
26	0.999999999999988\\
27	0.999999999999563\\
28	0.999999999988748\\
29	0.999999999791982\\
30	0.99999999715349\\
31	0.999999970376514\\
32	0.999999759628971\\
33	0.999998444021344\\
34	0.999991793270819\\
35	0.999964039200906\\
36	0.999866721326822\\
37	0.999575261839368\\
38	0.998818353603257\\
39	0.997090004274599\\
40	0.993575196663826\\
41	0.987134174200091\\
42	0.976382836996035\\
43	0.959874968653684\\
44	0.936350128082073\\
45	0.904981192982093\\
46	0.865552923170853\\
47	0.818526926292131\\
48	0.764985451221801\\
49	0.706479766412457\\
50	0.644827494811534\\
51	0.581905031048773\\
52	0.519470633618399\\
53	0.459038006125312\\
54	0.401805173068627\\
55	0.348632628098415\\
56	0.300058959867649\\
57	0.256340663167582\\
58	0.217504165744609\\
59	0.183400808106729\\
60	0.153758529275655\\
61	0.128226693510877\\
62	0.106412559954768\\
63	0.0879093080007558\\
64	0.072316374399702\\
65	0.0592532707728433\\
66	0.0483681698069731\\
67	0.0393424897360567\\
68	0.0318925549339405\\
69	0.0257692229160406\\
70	0.020756179835828\\
71	0.0166674365566811\\
72	0.0133444137189224\\
73	0.0106528885696271\\
74	0.00847998665385863\\
75	0.00673133423158456\\
76	0.00532843835596031\\
77	0.00420632706503787\\
78	0.0033114586479408\\
79	0.00259989363970248\\
80	0.00203571381869749\\
81	0.0015896672696043\\
82	0.00123801626121167\\
83	0.000961564262279224\\
84	0.000744839228789429\\
85	0.000575411814131566\\
86	0.000443329064182763\\
87	0.00034064622139296\\
88	0.000261041329294054\\
89	0.000199499302600104\\
90	0.000152053966097743\\
91	0.000115578230769131\\
92	8.76140488336175e-05\\
93	6.62351085010116e-05\\
94	4.99363525721108e-05\\
95	3.75453912192514e-05\\
96	2.81517125366726e-05\\
97	2.10503072749502e-05\\
98	1.5696922100239e-05\\
99	1.16726642348519e-05\\
100	8.65609173450466e-06\\
101	6.40128230711134e-06\\
102	4.720655247481e-06\\
103	3.47155959157242e-06\\
104	2.5458384755872e-06\\
105	1.86173460703376e-06\\
106	1.35763251696286e-06\\
107	9.87236003502368e-07\\
108	7.1586450201e-07\\
109	5.17617004813786e-07\\
110	3.73207898141104e-07\\
111	2.68320909935227e-07\\
112	1.92360612105702e-07\\
113	1.37509507225353e-07\\
114	9.80168076702981e-08\\
115	6.96653590370744e-08\\
116	4.93716344474038e-08\\
117	3.48883298917357e-08\\
118	2.4582163526965e-08\\
119	1.72701105638495e-08\\
120	1.20977005080292e-08\\
121	8.44960368340253e-09\\
122	5.88431703363312e-09\\
123	4.08579126087716e-09\\
124	2.82858536593267e-09\\
125	1.95245775103103e-09\\
126	1.34372157845064e-09\\
127	9.22000253922306e-10\\
128	6.30791419098387e-10\\
129	4.30247837357456e-10\\
130	2.92573076876579e-10\\
131	1.9838353182422e-10\\
132	1.34093625092646e-10\\
133	9.03810359886847e-11\\
134	6.0708771343343e-11\\
135	4.06430444854777e-11\\
136	2.71711542154662e-11\\
137	1.80762071977369e-11\\
138	1.19939613796305e-11\\
139	7.95807864051312e-12\\
140	5.22959453519434e-12\\
141	3.46744855050929e-12\\
142	2.27373675443232e-12\\
143	1.47792889038101e-12\\
144	9.66338120633736e-13\\
145	6.25277607468888e-13\\
146	3.97903932025656e-13\\
147	2.8421709430404e-13\\
148	1.70530256582424e-13\\
149	1.13686837721616e-13\\
150	5.6843418860808e-14\\
151	5.6843418860808e-14\\
152	0\\
153	0\\
154	0\\
155	0\\
156	0\\
157	0\\
158	0\\
159	0\\
160	0\\
161	0\\
162	0\\
163	0\\
164	0\\
165	0\\
166	0\\
167	0\\
168	0\\
169	0\\
170	0\\
171	0\\
172	0\\
173	0\\
174	0\\
175	0\\
176	0\\
177	0\\
178	0\\
179	0\\
180	0\\
181	0\\
182	0\\
183	0\\
184	0\\
185	0\\
186	0\\
187	0\\
188	0\\
189	0\\
190	0\\
191	0\\
192	0\\
193	0\\
194	0\\
195	0\\
196	0\\
197	0\\
198	0\\
199	0\\
200	0\\
};
\addplot [color=black, dashed, line width=0.9pt]
  table[row sep=crcr]{%
  40 0\\
  40 1\\
};
\addplot [color=black, dashed, line width=0.9pt]
  table[row sep=crcr]{%
  80 0\\
  80 1\\
};

\node at (rel axis cs:0.15,0.99) [anchor=north] {\textbf{\footnotesize{DANGEROUS}}};
\node at (rel axis cs:0.15,0.91) [anchor=north] {\textbf{\footnotesize{AREA}}};
\node at (rel axis cs:0.15,0.79) [anchor=north] {\footnotesize{Jamming}};
\node at (rel axis cs:0.15,0.72) [anchor=north] {\footnotesize{always successful}};

\node at (rel axis cs:0.5,0.99) [anchor=north] {\textbf{\footnotesize{CRITICAL}}};
\node at (rel axis cs:0.5,0.91) [anchor=north] {\textbf{\footnotesize{AREA}}};

\node at (rel axis cs:5/6,0.99) [anchor=north] {\textbf{\footnotesize{SAFE}}};
\node at (rel axis cs:5/6,0.91) [anchor=north] {\textbf{\footnotesize{AREA}}};
\node at (rel axis cs:5/6,0.79) [anchor=north] {\footnotesize{Jamming}};
\node at (rel axis cs:5/6,0.72) [anchor=north] {\footnotesize{always ineffective}};

\end{axis}
\end{tikzpicture}%

%% file: figures/Psucc_dist_ws.tex
%
%
\definecolor{mycolor1}{rgb}{0.92549,0.69020,0.12157}%
\definecolor{mycolor2}{rgb}{0.00000,0.56078,1.00000}%
\definecolor{mycolor3}{rgb}{0.60000,0.00000,0.60000}%
\begin{tikzpicture}
\pgfplotsset{every tick label/.append style={font=\footnotesize}}
\tikzstyle{dotted} = [dash pattern=on \pgflinewidth off 0.5mm] 
\tikzstyle{dashed} = [dash pattern=on 7.5*0.8*0.8pt off 7.5*0.4*0.8pt]
\tikzstyle{dashed2} = [dash pattern=on 7.5*0.5*0.8pt off 7.5*0.25*0.8pt]
\tikzstyle{dashdotted} = [dash pattern=on 7.5*0.8*0.6pt off 7.5*0.8*0.3pt on \the\pgflinewidth off 7.5*0.8*0.3pt]
\tikzstyle{dotted2} = [dash pattern=on 7.5*0.8*0.2pt off 7.5*0.8*0.1pt]

\begin{axis}[%
width=0.951\fwidth,
height=\fheight,
at={(0\fwidth,0\fheight)},
scale only axis,
xmin=1,
xmax=120,
xtick={1,20,40,60,80,100,120},
xticklabels={1,20,40,60,80,100,120},
xlabel style={font=\small\color{white!15!black}},
xlabel={Receiver-Jammer distance [m]},
ymin=0,
ymax=1,
ytick={0,0.1,0.2,0.3,0.4,0.5,0.6,0.7,0.8,0.9,1},
ylabel style={font=\small\color{white!15!black}},
ylabel={Success probability},
axis background/.style={fill=white},
xmajorgrids,
ymajorgrids,
legend style={font=\small, at={(0.98,0.02)}, anchor=south east, legend cell align=left, align=left, draw=white!15!black}
]
\addplot [color=red, line width=.9pt, mark size=2.2pt, mark=triangle*, mark options={solid}]
  table[row sep=crcr]{%
1	0.0988415303475288\\
20	0.0988415303475288\\
25	0.0988415303475288\\
30	0.0988415315865107\\
31	0.0988415424905822\\
35	0.0988559781599724\\
40	0.101180479014813\\
41	0.104370247809779\\
47	0.301806116445676\\
50	0.335553374245552\\
54	0.657859361748813\\
60	0.839194095660566\\
70	0.96514404581155\\
78	0.994341787741422\\
80	0.996517222464893\\
90	0.999739374037419\\
100	0.99998516100684\\
106	0.999997672612539\\
114	0.999999831969254\\
120	0.999999979260792\\
122	0.999999989912502\\
136	0.999999999953421\\
140	0.999999999991035\\
160	1\\
171	1\\
180	1\\
200	1\\
};
\addlegendentry{$\Gamma$ = $\infty$}

\addplot [color=mycolor1, line width=.9pt, mark size=2.2pt, mark=*, mark options={solid}]
  table[row sep=crcr]{%
1	0.2\\
20	0.2\\
25	0.2\\
30	0.326530612244898\\
31	0.428571428571429\\
35	0.419700020635817\\
40	0.418814223721296\\
41	0.39893319267362\\
47	0.431986481934587\\
50	0.479577597593174\\
54	0.901986416687844\\
60	0.976149316135049\\
70	0.96514404581155\\
78	0.994341787741422\\
80	0.996517222464893\\
90	0.999739374037419\\
100	0.99998516100684\\
106	0.999997672612539\\
114	0.999999831969254\\
120	0.999999979260792\\
122	0.999999989912502\\
136	0.999999999953421\\
140	0.999999999991035\\
160	1\\
171	1\\
180	1\\
200	1\\
};
\addlegendentry{$\Gamma$ = 1}

\addplot [color=mycolor2, line width=.9pt, mark size=2.9pt, mark=x, mark options={solid}]
  table[row sep=crcr]{%
1	0.142415768519163\\
20	0.142415768519163\\
25	0.128737878253397\\
30	0.0900302989977104\\
31	0.127103498741654\\
35	0.138569519856021\\
40	0.127994000788664\\
41	0.123425200495719\\
47	0.199979339170274\\
50	0.332042711257056\\
54	0.675134665946996\\
60	0.885400516781021\\
70	0.968589815883722\\
78	0.994340286845468\\
80	0.996516654817601\\
90	0.999739370866879\\
100	0.999985161001702\\
106	0.999997672612286\\
114	0.999999831969253\\
120	0.999999979260792\\
122	0.999999989912502\\
136	0.999999999953421\\
140	0.999999999991035\\
160	1\\
171	1\\
180	1\\
200	1\\
};
\addlegendentry{$\Gamma$ = 10}

\addplot [color=mycolor3, line width=.9pt, mark size=1.9pt, mark=square*, mark options={solid}]
  table[row sep=crcr]{%
1	0.0430663778853634\\
20	0.0430663778853634\\
25	0.0430663778853627\\
30	0.0430828135651608\\
31	0.0430825856377997\\
35	0.0430964837513927\\
40	0.0454083557812968\\
41	0.0481942677667726\\
47	0.138019172840467\\
50	0.274861099004791\\
54	0.517794978672595\\
60	0.78637116425496\\
70	0.965331488259548\\
78	0.994337281982521\\
80	0.996515518809545\\
90	0.999739363997086\\
100	0.999985160975156\\
106	0.999997672611781\\
114	0.99999983196925\\
120	0.999999979260792\\
122	0.999999989912502\\
136	0.999999999953421\\
140	0.999999999991035\\
160	1\\
171	1\\
180	1\\
200	1\\
};
\addlegendentry{$\Gamma$ = 30}

\end{axis}
\end{tikzpicture}%

%% file: figures/lifetime_distance_ws.tex
%
%
\definecolor{mycolor1}{rgb}{0.92549,0.69020,0.12157}%
\definecolor{mycolor2}{rgb}{0.00000,0.56078,1.00000}%
\definecolor{mycolor3}{rgb}{0.60000,0.00000,0.60000}%
\begin{tikzpicture}
\pgfplotsset{every tick label/.append style={font=\footnotesize}}
\tikzstyle{dotted} = [dash pattern=on \pgflinewidth off 0.5mm] 
\tikzstyle{dashed} = [dash pattern=on 7.5*0.8*0.8pt off 7.5*0.4*0.8pt]
\tikzstyle{dashed2} = [dash pattern=on 7.5*0.5*0.8pt off 7.5*0.25*0.8pt]
\tikzstyle{dashdotted} = [dash pattern=on 7.5*0.8*0.6pt off 7.5*0.8*0.3pt on \the\pgflinewidth off 7.5*0.8*0.3pt]
\tikzstyle{dotted2} = [dash pattern=on 7.5*0.8*0.2pt off 7.5*0.8*0.1pt]

\begin{axis}[%
width=0.951\fwidth,
height=\fheight,
at={(0\fwidth,0\fheight)},
scale only axis,
xmin=1,
xmax=120,
xtick={1,20,40,60,80,100,120},
xticklabels={1,20,40,60,80,100,120},
xlabel style={font=\small\color{white!15!black}},
xlabel={Receiver-Jammer distance [m]},
ymin=25,
ymax=50,
ytick={25,30,35,40,45,50},
ylabel style={font=\small\color{white!15!black}},
ylabel={Lifetime},
axis background/.style={fill=white},
xmajorgrids,
ymajorgrids,
legend style={font=\small, at={(0.98,0.02)}, anchor=south east, legend cell align=left, align=left, draw=white!15!black}
]
\addplot [color=red, line width=.9pt, mark size=2.2pt, mark=triangle*, mark options={solid}]
  table[row sep=crcr]{%
1	49.0043129122154\\
20	49.0043129122154\\
25	49.0043129122153\\
30	49.0043128983789\\
31	49.0043128623445\\
35	49.0042593411328\\
40	49.0216202099772\\
41	49.0109202921526\\
47	47.4771964245206\\
50	38.4832878327563\\
54	37.6011036564284\\
60	43.0895003496724\\
70	50\\
78	50\\
80	50\\
90	50\\
100	50\\
106	50\\
114	50\\
120	50\\
122	50\\
136	50\\
140	50\\
160	50\\
171	50\\
180	50\\
200	50\\
};
\addlegendentry{$\Gamma$ = $\infty$}

\addplot [color=mycolor1, line width=.9pt, mark size=2.2pt, mark=*, mark options={solid}]
  table[row sep=crcr]{%
1	50\\
20	50\\
25	50\\
30	49\\
31	49\\
35	48.2509064323288\\
40	48.1773534980743\\
41	46.5837409246301\\
47	49\\
50	25\\
54	25\\
60	35\\
70	50\\
78	50\\
80	50\\
90	50\\
100	50\\
106	50\\
114	50\\
120	50\\
122	50\\
136	50\\
140	50\\
160	50\\
171	50\\
180	50\\
200	50\\
};
\addlegendentry{$\Gamma$ = 1}

\addplot [color=mycolor2, line width=.9pt, mark size=2.9pt, mark=x, mark options={solid}]
  table[row sep=crcr]{%
1	43.2738520658975\\
20	43.2738520658975\\
25	43.8168486735615\\
30	44.1654753190764\\
31	42.8526371864454\\
35	43.9280636432967\\
40	43.549291910768\\
41	43.8130501603096\\
47	40.0119263632347\\
50	32.4855651908541\\
54	32.5591921344271\\
60	39.2784151746837\\
70	48.3065609510681\\
78	50\\
80	50\\
90	50\\
100	50\\
106	50\\
114	50\\
120	50\\
122	50\\
136	50\\
140	50\\
160	50\\
171	50\\
180	50\\
200	50\\
};
\addlegendentry{$\Gamma$ = 10}

\addplot [color=mycolor3, line width=.9pt, mark size=1.9pt, mark=square*, mark options={solid}]
  table[row sep=crcr]{%
1	48.9565333320481\\
20	48.9565333320481\\
25	48.9565333320482\\
30	48.9543302085752\\
31	48.9541789777156\\
35	48.9540363907754\\
40	48.9761902838468\\
41	48.9356687193564\\
47	47.4477124054023\\
50	44.943265982561\\
54	45.3758507694178\\
60	47.133880496659\\
70	49.0054856250666\\
78	50\\
80	50\\
90	50\\
100	50\\
106	50\\
114	50\\
120	50\\
122	50\\
136	50\\
140	50\\
160	50\\
171	50\\
180	50\\
200	50\\
};
\addlegendentry{$\Gamma$ = 30}

\end{axis}
\end{tikzpicture}%

%% file: figures/Psucc_dist_alpha.tex
%
%
\definecolor{mycolor1}{rgb}{0.92549,0.69020,0.12157}%
\definecolor{mycolor2}{rgb}{0.00000,0.56078,1.00000}%
\definecolor{mycolor3}{rgb}{0.60000,0.00000,0.60000}%
\begin{tikzpicture}
\pgfplotsset{every tick label/.append style={font=\footnotesize}}
\tikzstyle{dotted} = [dash pattern=on \pgflinewidth off 0.5mm] 
\tikzstyle{dashed} = [dash pattern=on 7.5*0.8*0.8pt off 7.5*0.4*0.8pt]
\tikzstyle{dashed2} = [dash pattern=on 7.5*0.5*0.8pt off 7.5*0.25*0.8pt]
\tikzstyle{dashdotted} = [dash pattern=on 7.5*0.8*0.6pt off 7.5*0.8*0.3pt on \the\pgflinewidth off 7.5*0.8*0.3pt]
\tikzstyle{dotted2} = [dash pattern=on 7.5*0.8*0.2pt off 7.5*0.8*0.1pt]

\begin{axis}[%
width=0.951\fwidth,
height=\fheight,
at={(0\fwidth,0\fheight)},
scale only axis,
xmin=1,
xmax=120,
xtick={1,20,40,60,80,100,120},
xticklabels={1,20,40,60,80,100,120},
xlabel style={font=\small\color{white!15!black}},
xlabel={Receiver-Jammer distance [m]},
ymin=0,
ymax=1,
ytick={0,0.1,0.2,0.3,0.4,0.5,0.6,0.7,0.8,0.9,1},
ylabel style={font=\small\color{white!15!black}},
ylabel={Success probability},
axis background/.style={fill=white},
xmajorgrids,
ymajorgrids,
legend style={font=\small, at={(0.98,0.02)}, anchor=south east, legend cell align=left, align=left, draw=white!15!black}
]
\addplot [color=red, line width=.9pt, mark size=2.2pt, mark=triangle*, mark options={solid}]
  table[row sep=crcr]{%
1	0.0473236071990899\\
20	0.0473236071990899\\
25	0.0473236071990897\\
30	0.047324629065167\\
31	0.047323599579797\\
35	0.0473379932008251\\
40	0.0499237120214676\\
41	0.0525899514151781\\
47	0.147916310318876\\
50	0.279829252460187\\
54	0.513245810355276\\
60	0.783059812466824\\
70	0.969019852333886\\
78	0.994418597029029\\
80	0.996515518809545\\
90	0.999739364525501\\
100	0.999985160976012\\
106	0.999997672611781\\
114	0.999999831969249\\
120	0.999999979260792\\
122	0.999999989912502\\
136	0.999999999953421\\
140	0.999999999991034\\
160	1\\
171	1\\
180	1\\
200	1\\
};
\addlegendentry{$\alpha$ = 0.2}

\addplot [color=mycolor1, line width=.9pt, mark size=2.2pt, mark=*, mark options={solid}]
  table[row sep=crcr]{%
1	0.0430663778853634\\
20	0.0430663778853634\\
25	0.0430663778853627\\
30	0.0430828135651608\\
31	0.0430825856377997\\
35	0.0430964837513927\\
40	0.0454083557812968\\
41	0.0481942677667726\\
47	0.138019172840467\\
50	0.274861099004791\\
54	0.517794978672595\\
60	0.78637116425496\\
70	0.965331488259548\\
78	0.994337281982521\\
80	0.996515518809545\\
90	0.999739363997086\\
100	0.999985160975156\\
106	0.999997672611781\\
114	0.99999983196925\\
120	0.999999979260792\\
122	0.999999989912502\\
136	0.999999999953421\\
140	0.999999999991035\\
160	1\\
171	1\\
180	1\\
200	1\\
};
\addlegendentry{$\alpha$ = 0.4}

\addplot [color=mycolor2, line width=.9pt, mark size=2.9pt, mark=x, mark options={solid}]
  table[row sep=crcr]{%
1	0.0611428571428572\\
20	0.0611428571428572\\
25	0.0634285714285715\\
30	0.0412303565913359\\
31	0.0385411682354798\\
35	0.0385801962637044\\
40	0.0408366289813897\\
41	0.0678147864301478\\
47	0.121460413961718\\
50	0.259681673863599\\
54	0.51409105798757\\
60	0.789863979911737\\
70	0.96526614506319\\
78	0.994337281982521\\
80	0.996515471515246\\
90	0.999739364525501\\
100	0.999985160975156\\
106	0.999997672611759\\
114	0.999999831969251\\
120	0.999999979260792\\
122	0.999999989912502\\
136	0.999999999953421\\
140	0.999999999991035\\
160	1\\
171	1\\
180	1\\
200	1\\
};
\addlegendentry{$\alpha$ = 0.6}

\addplot [color=mycolor3, line width=.9pt, mark size=1.9pt, mark=square*, mark options={solid}]
  table[row sep=crcr]{%
1	0.0285714285714285\\
20	0.0285714285714285\\
25	0.0285714285714287\\
30	0.0285714295473729\\
31	0.0285714387280324\\
35	0.0285837586315871\\
40	0.0307954742748699\\
41	0.033067935829751\\
47	0.144241852579245\\
50	0.247943777255433\\
54	0.455792802409903\\
60	0.760955505400742\\
70	0.964968245083236\\
78	0.994337281982521\\
80	0.996515518809545\\
90	0.999739364525501\\
100	0.999985160976012\\
106	0.999997672611575\\
114	0.999999831969252\\
120	0.999999979260792\\
122	0.999999989912502\\
136	0.999999999953421\\
140	0.999999999991035\\
160	1\\
171	1\\
180	1\\
200	1\\
};
\addlegendentry{$\alpha$ = 0.8}

\end{axis}
\end{tikzpicture}%

%% file: figures/lifetime_distance_alpha.tex
%
%
\definecolor{mycolor1}{rgb}{0.92549,0.69020,0.12157}%
\definecolor{mycolor2}{rgb}{0.00000,0.56078,1.00000}%
\definecolor{mycolor3}{rgb}{0.60000,0.00000,0.60000}%
\begin{tikzpicture}
\pgfplotsset{every tick label/.append style={font=\footnotesize}}
\tikzstyle{dotted} = [dash pattern=on \pgflinewidth off 0.5mm] 
\tikzstyle{dashed} = [dash pattern=on 7.5*0.8*0.8pt off 7.5*0.4*0.8pt]
\tikzstyle{dashed2} = [dash pattern=on 7.5*0.5*0.8pt off 7.5*0.25*0.8pt]
\tikzstyle{dashdotted} = [dash pattern=on 7.5*0.8*0.6pt off 7.5*0.8*0.3pt on \the\pgflinewidth off 7.5*0.8*0.3pt]
\tikzstyle{dotted2} = [dash pattern=on 7.5*0.8*0.2pt off 7.5*0.8*0.1pt]

\begin{axis}[%
width=0.951\fwidth,
height=\fheight,
at={(0\fwidth,0\fheight)},
scale only axis,
xmin=1,
xmax=120,
xtick={1,20,40,60,80,100,120},
xticklabels={1,20,40,60,80,100,120},
xlabel style={font=\small\color{white!15!black}},
xlabel={Receiver-Jammer distance [m]},
ymin=25,
ymax=50,
ytick={25,30,35,40,45,50},
ylabel style={font=\small\color{white!15!black}},
ylabel={Lifetime},
axis background/.style={fill=white},
xmajorgrids,
ymajorgrids,
legend style={font=\small, at={(0.98,0.02)}, anchor=south east, legend cell align=left, align=left, draw=white!15!black}
]
\addplot [color=red, line width=0.9pt, mark size=2.2pt, mark=triangle*, mark options={solid}]
  table[row sep=crcr]{%
1	48.3083717823545\\
20	48.3083717823545\\
25	48.3083717823544\\
30	48.3087716674534\\
31	48.3084283333498\\
35	48.3082989508229\\
40	48.2889236488738\\
41	48.2683092425239\\
47	46.3866892119589\\
50	43.8863774335663\\
54	44.6024431591329\\
60	46.88744852105\\
70	48.0238227290055\\
78	49.0037152974533\\
80	50\\
90	50\\
100	50\\
106	50\\
114	50\\
120	50\\
122	50\\
136	50\\
140	50\\
160	50\\
171	50\\
180	50\\
200	50\\
};
\addlegendentry{$\alpha$ = 0.2}

\addplot [color=mycolor1, line width=.9pt, mark size=2.2pt, mark=*, mark options={solid}]
  table[row sep=crcr]{%
1	48.9565333320481\\
20	48.9565333320481\\
25	48.9565333320482\\
30	48.9543302085752\\
31	48.9541789777156\\
35	48.9540363907754\\
40	48.9761902838468\\
41	48.9356687193564\\
47	47.4477124054023\\
50	44.943265982561\\
54	45.3758507694178\\
60	47.133880496659\\
70	49.0054856250666\\
78	50\\
80	50\\
90	50\\
100	50\\
106	50\\
114	50\\
120	50\\
122	50\\
136	50\\
140	50\\
160	50\\
171	50\\
180	50\\
200	50\\
};
\addlegendentry{$\alpha$ = 0.4}

\addplot [color=mycolor2, line width=.9pt, mark size=2.9pt, mark=x, mark options={solid}]
  table[row sep=crcr]{%
1	50\\
20	50\\
25	50\\
30	49.5000218862075\\
31	49.4161428488394\\
35	49.4152410034048\\
40	49.4082185772532\\
41	50\\
47	48.9085620053688\\
50	48.8769355361157\\
54	46.0805085710976\\
60	47.5981250800249\\
70	49.0471041959901\\
78	50\\
80	50\\
90	50\\
100	50\\
106	50\\
114	50\\
120	50\\
122	50\\
136	50\\
140	50\\
160	50\\
171	50\\
180	50\\
200	50\\
};
\addlegendentry{$\alpha$ = 0.6}

\addplot [color=mycolor3, line width=.9pt, mark size=1.9pt, mark=square*, mark options={solid}]
table[row sep=crcr]{%
1	50\\
20	50\\
25	50\\
30	50\\
31	50\\
35	50\\
40	50\\
41	50\\
47	50\\
50	50\\
54	50\\
60	49.0318907617943\\
70	50\\
78	50\\
80	50\\
90	50\\
100	50\\
106	50\\
114	50\\
120	50\\
122	50\\
136	50\\
140	50\\
160	50\\
171	50\\
180	50\\
200	50\\
};
\addlegendentry{$\alpha$ = 0.8}

\end{axis}
\end{tikzpicture}%

%% file: figures/Psucc_sensitivity.tex
%
%
\definecolor{mycolor1}{rgb}{0.92549,0.69020,0.12157}%
\definecolor{mycolor2}{rgb}{0.00000,0.56078,1.00000}%
\definecolor{mycolor3}{rgb}{0.60000,0.00000,0.60000}%
\definecolor{mycolor4}{rgb}{0.00000,0.49804,0.00000}%
\begin{tikzpicture}
\pgfplotsset{every tick label/.append style={font=\footnotesize}}
\tikzstyle{dotted} = [dash pattern=on \pgflinewidth off 0.5mm] 
\tikzstyle{dashed} = [dash pattern=on 7.5*0.8*0.8pt off 7.5*0.4*0.8pt]
\tikzstyle{dashed2} = [dash pattern=on 7.5*0.5*0.8pt off 7.5*0.25*0.8pt]
\tikzstyle{dashdotted} = [dash pattern=on 7.5*0.8*0.6pt off 7.5*0.8*0.3pt on \the\pgflinewidth off 7.5*0.8*0.3pt]
\tikzstyle{dotted2} = [dash pattern=on 7.5*0.8*0.2pt off 7.5*0.8*0.1pt]

\begin{axis}[%
width=0.951\fwidth,
height=\fheight,
at={(0\fwidth,0\fheight)},
scale only axis,
xmin=1,
xmax=120,
xtick={1,20,40,60,80,100,120},
xticklabels={1,20,40,60,80,100,120},
xlabel style={font=\small\color{white!15!black}},
xlabel={Receiver-Jammer distance [m]},
ymin=0,
ymax=1,
ytick={0,0.1,0.2,0.3,0.4,0.5,0.6,0.7,0.8,0.9,1},
ylabel style={font=\small\color{white!15!black}},
axis background/.style={fill=white},
xmajorgrids,
ymajorgrids,
legend style={font=\small, at={(0.02,0.98)}, anchor=north west, legend cell align=left, align=left, draw=white!15!black}
]
\addplot [color=red, line width=.9pt, mark size=2.2pt, mark=triangle*, mark options={solid}]
  table[row sep=crcr]{%
1	0.0428631685514169\\
20	0.0426927920104212\\
25	0.0431553770444348\\
30	0.0476711933709654\\
31	0.0457095419018671\\
35	0.0417736125532544\\
40	0.0500161720377327\\
41	0.048240660732378\\
47	0.136395745720148\\
50	0.267501886428774\\
54	0.520715069286762\\
60	0.79154024051804\\
70	0.972653061224489\\
78	0.9938\\
80	0.9968\\
90	1\\
100	1\\
106	1\\
114	1\\
120	1\\
122	1\\
136	1\\
140	1\\
160	1\\
171	1\\
180	1\\
200	1\\
};
\addlegendentry{error $\sigma$ = 0}

\addplot [color=mycolor1, line width=.9pt, mark size=2.2pt, mark=*, mark options={solid}]
  table[row sep=crcr]{%
1	0.0450680814879143\\
20	0.0452307101008766\\
25	0.0501315421913446\\
30	0.0495163844261109\\
31	0.05184726443769\\
35	0.052793968012737\\
40	0.0496975603654968\\
41	0.0531631965552178\\
47	0.13219456584669\\
50	0.264790875864903\\
54	0.496548580158663\\
60	0.763199390995992\\
70	0.89265306122449\\
78	0.9028\\
80	0.8902\\
90	0.895\\
100	0.8926\\
106	0.9176\\
114	0.9122\\
120	0.9064\\
122	0.889\\
136	0.889999999999999\\
140	0.906999999999999\\
160	0.9222\\
171	0.9198\\
180	0.9298\\
200	0.9288\\
};
\addlegendentry{error $\sigma$ = 0.05}

\addplot [color=mycolor2, line width=.9pt, mark size=2.9pt, mark=x, mark options={solid}]
  table[row sep=crcr]{%
1	0.0494977622131183\\
20	0.0564943262411347\\
25	0.0554043638732088\\
30	0.0606324405155216\\
31	0.050385565928499\\
35	0.0509656245476914\\
40	0.0616735801443611\\
41	0.0628301309885656\\
47	0.127530961243579\\
50	0.25777289546716\\
54	0.500409399265551\\
60	0.7524427613321\\
70	0.800816326530612\\
78	0.828\\
80	0.8328\\
90	0.843\\
100	0.8246\\
106	0.8104\\
114	0.8352\\
120	0.8084\\
122	0.799\\
136	0.8258\\
140	0.8348\\
160	0.8772\\
171	0.872\\
180	0.845999999999999\\
200	0.8756\\
};
\addlegendentry{error $\sigma$ = 0.1}

\addplot [color=mycolor3, line width=.9pt, mark size=1.9pt, mark=square*, mark options={solid}]
  table[row sep=crcr]{%
1	0.0584454371110146\\
20	0.0569683094514401\\
25	0.0597062383973015\\
30	0.0595604646113764\\
31	0.0638760312635692\\
35	0.0492013363162101\\
40	0.0590521819366044\\
41	0.0621490909777417\\
47	0.140716658957755\\
50	0.278721425254416\\
54	0.447091324904924\\
60	0.713244295405489\\
70	0.771632653061224\\
78	0.7708\\
80	0.7356\\
90	0.746\\
100	0.7714\\
106	0.751\\
114	0.7572\\
120	0.747\\
122	0.7536\\
136	0.7708\\
140	0.7612\\
160	0.859\\
171	0.7886\\
180	0.8188\\
200	0.8392\\
};
\addlegendentry{error $\sigma$ = 0.15}

\addplot [color=mycolor4, line width=.9pt, mark size=2.1pt, mark=diamond*, mark options={solid}]
  table[row sep=crcr]{%
1	0.0595300507529561\\
20	0.071159759009987\\
25	0.060518041684759\\
30	0.0550762866015971\\
31	0.071831891452233\\
35	0.0626317918656824\\
40	0.0653391156462585\\
41	0.0714455041313472\\
47	0.152417396653388\\
50	0.275480014416045\\
54	0.469307978022594\\
60	0.624434472678988\\
70	0.669183673469388\\
78	0.7178\\
80	0.7242\\
90	0.6792\\
100	0.708\\
106	0.6762\\
114	0.7078\\
120	0.7106\\
122	0.6772\\
136	0.7322\\
140	0.745\\
160	0.8072\\
171	0.8058\\
180	0.7408\\
200	0.7552\\
};
\addlegendentry{error $\sigma$ = 0.2}

\end{axis}
\end{tikzpicture}%

%% file: figures/PER_seaTrial.tex
%
%
\definecolor{mycolor1}{rgb}{0.92549,0.69020,0.12157}%
\definecolor{mycolor2}{rgb}{0.00000,0.56078,1.00000}%
\begin{tikzpicture}
\pgfplotsset{every tick label/.append style={font=\footnotesize}}
\tikzstyle{dotted} = [dash pattern=on \pgflinewidth off 0.5mm] 
\tikzstyle{dashed} = [dash pattern=on 7.5*0.8*0.8pt off 7.5*0.4*0.8pt]
\tikzstyle{dashed2} = [dash pattern=on 7.5*0.5*0.8pt off 7.5*0.25*0.8pt]
\tikzstyle{dashdotted} = [dash pattern=on 7.5*0.8*0.6pt off 7.5*0.8*0.3pt on \the\pgflinewidth off 7.5*0.8*0.3pt]
\tikzstyle{dotted2} = [dash pattern=on 7.5*0.8*0.2pt off 7.5*0.8*0.1pt]

\begin{axis}[%
width=0.951\fwidth,
height=\fheight,
at={(0\fwidth,0\fheight)},
scale only axis,
xmin=20,
xmax=171,
xlabel style={font=\small\color{white!15!black}},
xlabel={Receiver-Jammer distance [m]},
ymin=0,
ymax=1,
ytick={0,0.1,0.2,0.3,0.4,0.5,0.6,0.7,0.8,0.9,1},
ylabel style={font=\small\color{white!15!black}},
ylabel={Packet error rate},
axis background/.style={fill=white},
xmajorgrids,
ymajorgrids,
legend style={font=\small, at={(0.98,0.98)}, anchor=north east, legend cell align=left, align=left, draw=white!15!black}
]
\addplot [color=red, line width=.9pt, mark size=2.2pt, mark=triangle*, mark options={solid}]
  table[row sep=crcr]{%
31	0.9375\\
41	0.820833333333333\\
47	0.661538461538462\\
54	0.479166666666667\\
78	0.395833333333333\\
106	0.358333333333333\\
114	0.229166666666667\\
122	0.0916666666666667\\
136	0.104166666666667\\
171	0.0375\\
};
\addlegendentry{Lake test}

\addplot [color=mycolor1, line width=.9pt, mark size=2.2pt, mark=*, mark options={solid}]
  table[row sep=crcr]{%
1	1\\
20	1\\
25	1\\
30	0.99999999715349\\
31	0.999999970376514\\
35	0.999964039200906\\
40	0.993575196663826\\
41	0.987134174200091\\
47	0.818526926292131\\
50	0.644827494811534\\
54	0.401805173068627\\
60	0.153758529275655\\
70	0.020756179835828\\
78	0.0033114586479408\\
80	0.00203571381869749\\
90	0.000152053966097743\\
100	8.65609173450466e-06\\
106	1.35763251696286e-06\\
114	9.80168076702981e-08\\
120	1.20977005080292e-08\\
122	5.88431703363312e-09\\
136	2.71711542154662e-11\\
140	5.22959453519434e-12\\
160	0\\
171	0\\
180	0\\
200	0\\
};
\addlegendentry{LOS Uncoded}

\addplot [color=mycolor2, line width=.9pt, mark size=2.9pt, mark=x, mark options={solid}]
  table[row sep=crcr]{%
1	1\\
20	1\\
25	0.999999035425354\\
30	0.874323161360996\\
31	0.685649306237761\\
35	0.0228643169410362\\
40	9.74299683015579e-07\\
41	6.7750434949525e-08\\
47	2.43584743842609e-16\\
50	2.49709579373881e-21\\
54	1.36333539675284e-28\\
60	1.43279993941337e-40\\
70	5.36613397213873e-63\\
78	5.21661213963948e-83\\
80	2.6836262400025e-88\\
90	1.78546235757058e-116\\
100	1.36197390578112e-147\\
106	1.0486812528494e-167\\
114	3.04505777442999e-196\\
120	5.87018833330972e-219\\
122	8.77976323682035e-227\\
136	3.57917889309979e-285\\
140	0\\
160	0\\
171	0\\
180	0\\
200	0\\
};
\addlegendentry{LOS Coded}

\end{axis}
\end{tikzpicture}%

%% file: figures/Psucc_dist_seaTrial.tex
%
%
\definecolor{mycolor1}{rgb}{0.92549,0.69020,0.12157}%
\definecolor{mycolor2}{rgb}{0.00000,0.56078,1.00000}%
\definecolor{mycolor3}{rgb}{0.60000,0.00000,0.60000}%
\begin{tikzpicture}
\pgfplotsset{every tick label/.append style={font=\footnotesize}}
\tikzstyle{dotted} = [dash pattern=on \pgflinewidth off 0.5mm] 
\tikzstyle{dashed} = [dash pattern=on 7.5*0.8*0.8pt off 7.5*0.4*0.8pt]
\tikzstyle{dashed2} = [dash pattern=on 7.5*0.5*0.8pt off 7.5*0.25*0.8pt]
\tikzstyle{dashdotted} = [dash pattern=on 7.5*0.8*0.6pt off 7.5*0.8*0.3pt on \the\pgflinewidth off 7.5*0.8*0.3pt]
\tikzstyle{dotted2} = [dash pattern=on 7.5*0.8*0.2pt off 7.5*0.8*0.1pt]

\begin{axis}[%
width=0.951\fwidth,
height=\fheight,
at={(0\fwidth,0\fheight)},
scale only axis,
xmin=31,
xmax=171,
xlabel style={font=\small\color{white!15!black}},
xlabel={Receiver-Jammer distance [m]},
ymin=0,
ymax=1,
ytick={0,0.1,0.2,0.3,0.4,0.5,0.6,0.7,0.8,0.9,1},
ylabel style={font=\small\color{white!15!black}},
ylabel={Success probability},
axis background/.style={fill=white},
xmajorgrids,
ymajorgrids,
legend style={font=\small, at={(0.98,0.02)}, anchor=south east, legend cell align=left, align=left, draw=white!15!black}
]
\addplot [color=red, line width=.9pt, mark size=2.2pt, mark=triangle*, mark options={solid}]
  table[row sep=crcr]{%
31	0.0597002125142379\\
41	0.137568656929315\\
47	0.239468501198238\\
54	0.423102090674184\\
78	0.49935463941743\\
106	0.538114229249011\\
114	0.665059903381643\\
122	0.823900483091787\\
136	0.805608695652175\\
171	0.897826086956519\\
};
\addlegendentry{Lake test strategy}

\addplot [color=mycolor1, line width=.9pt, mark size=2.2pt, mark=*, mark options={solid}]
  table[row sep=crcr]{%
31	0.0572424645862044\\
41	0.110644559018797\\
47	0.213195257965827\\
54	0.402585046580514\\
78	0.44404\\
106	0.47608\\
114	0.60208\\
122	0.77304\\
136	0.75784\\
171	0.846559999999998\\
};
\addlegendentry{Uncoded strategy}

\addplot [color=mycolor2, line width=.9pt, mark size=2.9pt, mark=x, mark options={solid}]
  table[row sep=crcr]{%
31	0.0718837817492657\\
41	0.11644\\
47	0.832239999999998\\
54	0.78656\\
78	0.800039999999999\\
106	0.799759999999998\\
114	0.820679999999998\\
122	0.840599999999998\\
136	0.835599999999998\\
171	0.849199999999999\\
};
\addlegendentry{Coded strategy}

\addplot [color=mycolor3, line width=.9pt, mark size=1.9pt, mark=square*, mark options={solid}]
  table[row sep=crcr]{%
31	0.164625\\
41	0.175\\
47	0.2953\\
54	0.457380952380952\\
78	0.538418604651162\\
106	0.577813953488372\\
114	0.700266666666665\\
122	0.849866666666665\\
136	0.83422222222222\\
171	0.898260869565216\\
};
\addlegendentry{Dummy strategy}

\end{axis}
\end{tikzpicture}%

%% file: figures/lifetime_distance_seaTrial.tex
%
%
\definecolor{mycolor1}{rgb}{0.92549,0.69020,0.12157}%
\definecolor{mycolor2}{rgb}{0.00000,0.56078,1.00000}%
\definecolor{mycolor3}{rgb}{0.60000,0.00000,0.60000}%
\begin{tikzpicture}
\pgfplotsset{every tick label/.append style={font=\footnotesize}}
\tikzstyle{dotted} = [dash pattern=on \pgflinewidth off 0.5mm] 
\tikzstyle{dashed} = [dash pattern=on 7.5*0.8*0.8pt off 7.5*0.4*0.8pt]
\tikzstyle{dashed2} = [dash pattern=on 7.5*0.5*0.8pt off 7.5*0.25*0.8pt]
\tikzstyle{dashdotted} = [dash pattern=on 7.5*0.8*0.6pt off 7.5*0.8*0.3pt on \the\pgflinewidth off 7.5*0.8*0.3pt]
\tikzstyle{dotted2} = [dash pattern=on 7.5*0.8*0.2pt off 7.5*0.8*0.1pt]

\begin{axis}[%
width=0.951\fwidth,
height=\fheight,
at={(0\fwidth,0\fheight)},
scale only axis,
xmin=31,
xmax=171,
xlabel style={font=\small\color{white!15!black}},
xlabel={Receiver-Jammer distance [m]},
ymin=25,
ymax=50,
ytick={25,30,35,40,45,50},
ylabel style={font=\small\color{white!15!black}},
ylabel={Lifetime},
axis background/.style={fill=white},
xmajorgrids,
ymajorgrids,
legend style={font=\small, at={(0.98,0.02)}, anchor=south east, legend cell align=left, align=left, draw=white!15!black}
]
\addplot [color=red, line width=.9pt, mark size=2.2pt, mark=triangle*, mark options={solid}]
  table[row sep=crcr]{%
31	48.588\\
41	48.15\\
47	46.06\\
54	43.682\\
78	44.304\\
106	44.43\\
114	45.126\\
122	45.998\\
136	46\\
171	46\\
};
\addlegendentry{Lake test strategy}

\addplot [color=mycolor1, line width=.9pt, mark size=2.2pt, mark=*, mark options={solid}]
  table[row sep=crcr]{%
31	48.984\\
41	49.03\\
47	47.522\\
54	45.396\\
78	50\\
106	50\\
114	50\\
122	50\\
136	50\\
171	50\\
};
\addlegendentry{Uncoded strategy}

\addplot [color=mycolor2, line width=.9pt, mark size=2.9pt, mark=x, mark options={solid}]
  table[row sep=crcr]{%
31	46.214\\
41	50\\
47	50\\
54	50\\
78	50\\
106	50\\
114	50\\
122	50\\
136	50\\
171	50\\
};
\addlegendentry{Coded strategy}

\addplot [color=mycolor3, line width=.9pt, mark size=1.9pt, mark=square*, mark options={solid}]
  table[row sep=crcr]{%
31	48\\
41	40\\
47	40\\
54	42\\
78	43\\
106	43\\
114	45\\
122	45\\
136	45\\
171	46\\
};
\addlegendentry{Dummy strategy}
\end{axis}
\end{tikzpicture}%